\documentclass[aps,amsmath,amsfonts,12pt]{revtex4}
\usepackage{graphicx}
\usepackage{epstopdf}
\usepackage{epsfig,graphicx}
\usepackage[english]{babel}
\usepackage{amsfonts}
\usepackage{amsmath}
\usepackage{latexsym}
\usepackage{graphics,bm}
\usepackage{dcolumn}
\usepackage{natbib}
\usepackage{bm}
\usepackage{rotating}

\begin{document}

\title{Quantum Optical Response of a Hybrid Optomechanical Device embedded with a Qubit}

\author{ Sabur A. Barbhuiya and Aranya B Bhattacherjee }

\address{Department of Physics, Birla Institute of Technology and Science, Pilani,
Hyderabad Campus,  Telangana - 500078, India}

\begin{abstract}
We theoretically investigate the optical response in a hybrid quantum optomechanical system consisting of two optically coupled micro-cavities in which a two-level system (qubit) is embedded on a movable membrane. The qubit can either be a defect which interacts with the mechanical oscillator via the linear Jaynes-Cummings interaction or a superconducting charge qubit coupled with the mechanical mode via nonlinear interaction. We find that coherent perfect transmission (CPT), coherent perfect synthesis (CPS) and optomechanically induced absorption (OMIA) can be generated by suitably adjusting the system parameters. We find that the qubit and its interaction with the mechanical oscillator emerges as a new handle to control these quantum optical properties. The presence of the qubit results in four points where CPT and CPS can be realized compared to the pure optomechanical case (i.e. in the absence of qubit) where only three points are attained. This shows that the presence of the qubit gives us more flexibility in choosing the appropriate parameter regime where CPT and CPS can be attained and controlled. We also find that OMIA shows three distinct peaks both in the linear and nonlinear cases. In the absence of the qubit, OMIA is converted to optomechanically induced transmission (OMIT). An increase in in the qubit decay rate also shows a transition from OMIA to OMIT. Our study reveals that the optical response of the nonlinear case is relatively rapid (more sensitive) compared to the linear case to changes in the system parameters. This demonstrates the potential use of this hybrid system in designing tunable all-optical-switch and photon-router both of which forms an important element of a quantum information network.
\end{abstract}

\pacs{03.75.Kk,03.75.Lm, 42.50.Lc, 03.65.Ta, 05.40.Jc, 04.80.Nn}

\maketitle

\section{Introduction}

Since the past few decades there has been tremendous advancement in the understanding of light-matter interaction in hybrid optomechanical systems \citep{1}. In recent years research in the area of micro and nanoscale mechanical resonators have opened up the possibility of novel quantum devices \citep{2,3}. New and interesting physics have emerged by coupling of mechanical resonators with other quantum objects such as Superconducting charge qubits \citep{4,5,6,7,8,9,10,11,12,13,14,15}, transmission line resonators \citep{16,17,18,19,20}, optical cavities \citep{21,22,23,24}, quantum dots, nitrogen-vacancy centres(NV) \citep{25,26,27} and electron spin \citep{28}. As a result, importance of research has enhanced in the field of designing classical and quantum information processing systems using hybrid optomechanics \citep{29,30,31,32,33,34}. Experimental results have demonstrated that mechanical resonators which can be operated in quantum regimes \citep{36,37,38}, can be used as switches, data buses \citep{39} or transducers \citep{40,41}.  In recent years, experiments with cavity optomechanics have successfully entered into the resolved sideband limit where mechanical side-bands of the optical mode lie outside its linewidth \citep{42}. It has been shown that the intracavity optical field can modify the effective loss factor of mechanical mode \citep{30} which leads to mechanical cooling\citep{31} via the optomechanical interaction when the input field is red-detuned from the cavity resonance, where photons preferentially absorb a phonon from the mechanical oscillator and scatter upwards to the cavity resonance. This situation is quite similar to laser cooling of atomic and molecular motion in a cavity \citep{45}.

Hybrid electro-optomechanical systems demonstrate strong Kerr nonlinearities even in the weak-coupling regime \citep{46}, which can be used for photon blockade and generation of nonclassical states of microwave radiation. A double optomechanical cavity has also been shown to act as an optical switch by controlling the probe photon transmission \citep{47,48,49,50,51,52}. An interesting and useful development took place when a strong single photon optomechanical coupling was demonstrated \citep{53,54,55,56}. Electromagnetically induced transparency (EIT) has played a crucial role in many subfields of quantum optics. The quantum interference in the phonon excitation pathways leads to the optomechanical analog of EIT, the so called optomechanically induced transparency (OMIT) \citep{57,58,59}.

The OMIT phenomenon can be used to slow \citep{60,61,62} and even stop light signals \citep{60,61} which can be used to store information in mechanical oscillators. Multimode optomechanical systems have also been studied to attain quantum entanglement \citep{65}, OMIT \citep{66} and single photon nonlinearity \citep{67}.

In this paper, we investigate a double cavity optomechanical system with movable mirror in the middle, in the presence of linear/non-linear interaction of a two level system and the mechanical mode of the movable mirror. The linear interaction is acheived by embedding a quantum dot on the movable middle mirror while the nonlinear interaction is generated by coupling a superconducting qubit with nanomechanical oscillator \citep{69}. We have compared the two cases in terms of optical response (Coherent perfect transmission, coherent perfect synthesis, Electromagnetically induced transparency) of the system.

\section{The Physical Models}

We consider a hybrid double-cavity optomechanical system composed of two fixed mirrors with partial transmissivity and one movable mirror located between the two fixed mirrors (membrane in the middle) as shown in fig.1 \citep{70,71}. The membrane in the middle oscillator has an eigen frequency $\omega_{m}$ and a decay rate $\gamma_{m}$. The movable membrane is perfectly reflective and is at its equilibrium position so that the system can be regarded as two identical Febry-Perot cavities of length L and frequency $\omega_{0}$. The left(right) cavity optical mode is described by creation operator ${c_{1}}^{\dagger} ({c_{2}}^{\dagger})$ and annihilation operator $c_{1} (c_{2})$ while the mechanical mode is described by the creation and annihilation operator $b^{\dagger}$ and $b$ respectively. These operators satisfy the bosonic commutation relation $[A_{i}, {A_{j}}^{\dagger}]=\delta_{ij}$ $ ( A=c_{1}, c_{2}, b $ and $ i,j=1,2)$.

\begin{figure}[ht]
\hspace{-0.0cm}
\begin{tabular}{cc}
\includegraphics [scale=0.5]{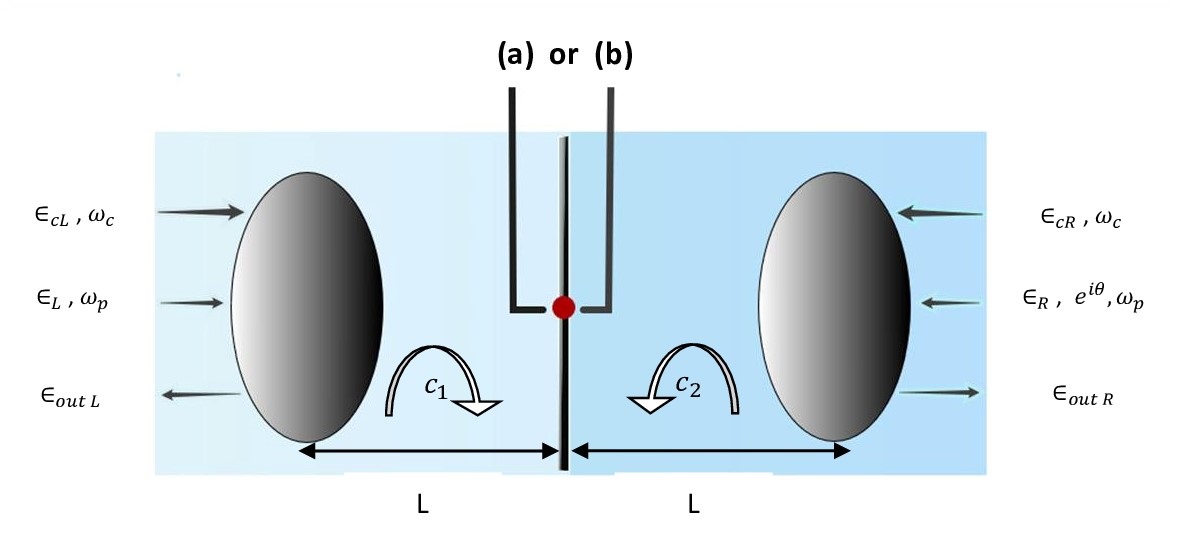}\\
\includegraphics [scale=2.5] {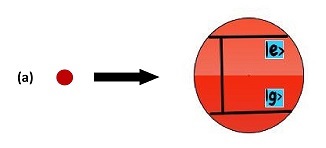}     \includegraphics [scale=2.5] {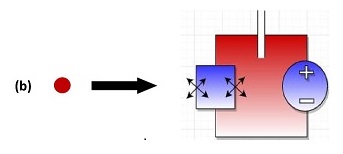} \\
 \end{tabular}
\caption{(Color online) Schematic diagram of the hybrid optomechanical system consisting of a double cavity with a semi-transparent movable mirror in the middle which corresponds to a mechanical oscillator (as shown in Fig.1). The oscillating mirror has a two-level system which could be simply a defect (Fig.1a) or a superconducting charge qubit (Fig.1b).  The two-level system interacts with the mechanical mode via linear coupling in case of a defect or nonlinear coupling for the superconducting qubit. In addition, the mechanical oscillator couples to the two cavity modes via radiation pressure. Both the cavities are driven by two separate control fields as shown and two separate probe fields are also incident on the two cavities from either side which have a phase difference of $\theta$.  }
\label{modfig}
\end{figure}

The system is driven from left and right fixed mirrors by two control(probe) fields with amplitudes, ${\epsilon_{cL}}=\sqrt{\frac{{2} k P_{cL} } { \hbar \omega_{c} }}$ and  ${\epsilon_{cR}}= \sqrt{\frac{{2} k P_{cR} }{\hbar \omega_{c} }}$ ( ${\epsilon_{L}}=\sqrt{{\frac{{2}{k}{P_{L}}}{{\hbar}\omega_{p}}}}$ and ${\epsilon_{R}}=\sqrt{{\frac{{2}{k}{P_{R}}}{{\hbar}\omega_{p}}}}$ ) respectively. 
Here the subscripts L(R) denotes the left (right) cavity. We assume both the cavities have the same decay rates $\kappa$. Both the left and right control (probe) modes have the same frequency $\omega_{c} (\omega_{p})$. Here  $P_{cL}$ , $P_{cR}$, $P_{L}$, and $P_{R}$ are the relevant field powers. 

We now discuss the two cases that we will be analyzing in the paper. The first case is that of a two-level system(qubit) linearly coupled to the mechanical oscillator. The qubit could be an intrinsic defect inside the mechanical resonator, a quantum dot or another two level system. The mechanical oscillator is coupled to the qubit via the linear Jaynes-Cummings interaction. The radiation-pressure Hamiltonian describes the interaction between the cavity modes and the mechanical mode. However there is no direct interaction between the qubit and the optical modes. Thus the total Hamiltonian is the frame rotating with respect to the control field frequency $\omega_{c}$ can be written as, 

\begin{equation}
H_{total}=H_{1,2} + H_{probe}+ H_{qd-m},
\end{equation}

\begin{equation}
H_{1,2}={\hbar}\Delta_{c} (c_{1}^{\dagger}{c_{1}}+c_{2}^{\dagger}{c_{2}})+{\hbar}{g_{0}}(c_{2}^{\dagger}{c_{2}}-c_{1}^{\dagger}{c_{1}})(b^{\dagger}+b)+{\hbar}{\omega_{m}}{b}^{\dagger}{b}+ i {\hbar}\epsilon_{cL} (c_{1}^{\dagger}-c_{1}) + i  {\hbar} \epsilon_{cR} (c_{2}^{\dagger}-c_{2}),
\end{equation}

\begin{equation}
H_{probe}=i{\hbar}(c_{1}^{\dagger} \epsilon_{L} e^{-i{\delta}t}-c_{1}{\epsilon_{L}}^{*}{e^{i{\delta}t}})+ i{\hbar}(c_{2}^{\dagger}{\epsilon_{R}}{e^{i{\theta}}}{e^{-i{\delta}t}}- c_{2}{\epsilon_{R}}^{*}{e^{-i{\theta}}}{e^{i{\delta}t}}),
\end{equation}

\begin{equation}
H_{qd-m}={\frac{1}{2}}\omega_{q}{\sigma_{z}}+{\hbar}g(b{\sigma}^{+}+b^{\dagger}{\sigma}^{-}).
\end{equation}

Here, ${\Delta_{c}}={\omega_{0}}-{\omega_{c}}$ denotes the detuning between the cavity mode and the control field, ${g_{0}}=\frac{\omega_{0}}{L} (\sqrt{\frac{\hbar}{2 m {\omega_{m}}}})$ is the optomechanical coupling constant, ${\delta}={\omega_{p}}-{\omega_{c}}$ is the detuning between the probe and the coupling field and $\theta$ is the relative phase between the left and right probe fields. Also, $\omega_{q}$ is the transition frequency of the two level system while $\sigma_{+}$, $\sigma_{-}$ and $\sigma_{z}$ are the usual Pauli operators describing the two-level system. The parameter $g$ describes the linear coupling strength between the mechanical resonator and the qubit. The linear system has been studied earlier in the case of EIT \citep{73}.

We now describe the optomechanical system in which the qubit is interacting non linearly with the mechanical mode \citep{74}. The nonlinear interaction can be acheived by embedding a superconducting charge qubit in the movable membrane in the middle of the cavity. As a result, the qubit-mechanical mode term is written as 

\begin{equation}
H_{qubit-m}= \frac{1}{2} \omega_{q} \sigma_{z}+{\hbar} g_{N} ( {b}^{2}{\sigma}^{+}+{b^{\dagger}}^{2}{\sigma}^{-}).
\end{equation}

Here $g_{N}$ is the coupling strength between the superconducting qubit and the mechanical oscillator. The origin of $g_{N}$ is the Josephson coupling energy in a Cooper-pair box.

\section{Heisenberg-Langevin equations, steady state and fluctuation dynamics}

We now proceed ahead to study the quantum dynamics of the linear and non-linear case systematically.

\subsection{Linear Case}

Considering relevant dissipation and quantum or thermal noise, the quantum dynamics of the total system's operators is given by the following quantum-Langevin equations

\begin{equation}
\dot{b}= -i {\omega_{m}}b(t) - i {g_{0}}(c_{2}^{\dagger}{c_{2}}-c_{1}^{\dagger}{c_{1}})-i g {\sigma}^{-}(t) - {\frac{\gamma_{m}}{2}}b(t) + \sqrt{\gamma_{m}} b_{in},
\end{equation}

\begin{equation}
\dot{\sigma}^{-}= -i {\omega_{q}}{\sigma}^{-}(t) +i g b {\sigma}_{z}(t) - {\frac{k_{d}}{2}}{\sigma}^{-}(t)+ \sqrt k_{d}{{\sigma}^{-}_{in}},
\end{equation}

\begin{equation}
\dot{c_{1}}= -[k+ i {\Delta_{c}} -  i {g_{0}}(b^{\dagger}+b)] c_{1} + {\epsilon_{cL}} + {\epsilon_{L}} {e^{-i{\delta}t}}+ \sqrt{2k} {c_{1}}^{in},
\end{equation}

\begin{equation}
\dot{c_{2}}= -[k+ i {\Delta_{c}} +  i {g_{0}}(b^{\dagger}+b)] c_{2} + {\epsilon_{cR}} + {\epsilon_{R}}{e^{i{\theta}}}{e^{-i{\delta}t}}+ \sqrt{2k} {c_{2}}^{in}.
\end{equation}

Here $b_{in}$, ${\sigma}^{-}_{in}$ are the zero-mean-value environmental noise operators of the mechanical oscillator and the two-level qubit respectively. Also ${c_{1}}^{in}({c_{2}}^{in})$ is the zero-mean-value quantum noise operators of the left(right) cavity. The probe fields are small compared to the control and hence can be considered as comparable to noise.

In the classical limit, we drop the fluctuations, the probe fields and replace the operators by their expectation values. We can generate the steady-state mean values by setting all the time derivative to zero and with the factorization assumption $<b c_{i}>=<b><c_{i}>$ 

\begin{equation}
\langle b \rangle=b_{s}=\frac{- i g_{0} ( {|c_{2s}|}^2-{|c_{1s}|}^2) (\frac{k_{d}}{2}+i \omega_{q}) } { (\frac{\gamma_{m}}{2}+i \omega_{m})(\frac{k_{d}}{2}+i \omega_{q}) -g^{2} { \langle \sigma_{z} \rangle}_{s}},
\end{equation}

\begin{equation}
\langle {\sigma^{-}} \rangle= {{\sigma}_{s}}^{-}= \frac{ i g b_{s} \langle {{\sigma_{z}} \rangle}_{s}}{({\frac{k_{d}}{2}}+i {\omega_{q}})},
\end{equation}

\begin{equation}
\langle {c_{1}} \rangle=c_{1s}=\frac{\epsilon_{cL}}{k+i{\Delta_{1}}},
\end{equation}

\begin{equation}
\langle {c_{2}} \rangle=c_{2s}=\frac{\epsilon_{cR}}{k+i{\Delta_{2}}}.
\end{equation}

Where, $\Delta_{1,2}=\Delta_{c} {\mp} g_{0} (b_{s} + {b_{s}}^{*}) $ is the effective detuning of the cavity modes. Note that the term $g_{0} (b_{s} + {b_{s}}^{*}) \leq \Delta_{c}$, when $g_{0}$ is weak ($g_{0} \leq \omega_{m}$) and the number of photons in the two cavities are same. This is evident from the expression for $b_{s}$ since if  ${|c_{1s}|}^2 \approx {|c_{2s}|}^2$ then $b_{s}\ll 1$ . 

We now derive the quantum Langevin equations by substituting the ansatz $b= b_{s} + {\delta} b$ , $c_{1}= c_{1s} + {\delta} c_{1}$ , $c_{2}= c_{2s} + {\delta} c_{2}$ and ${\sigma}^{-}= {{\sigma}^{-}}_{s} + {\delta}{\sigma}^{-}$ into eqns.(6)-(9) and retain only the first order terms in the fluctuations $\delta{b}$, $\delta{c_{1}}$, $\delta{c_{2}}$ and ${\delta}{\sigma}^{-}$. We are essentially obtaining the linearized quantum-Langevin equations for the fluctuations. We assume that each control field drives the corresponding cavity mode at the mechanical red sideband $( \Delta_{1} \approx \Delta_{2} \approx \omega_{m} \approx \omega_{q})$ and simultaneously, the optomechanical system is operated in the resolved sideband regime $(\omega_{m} \gg k)$. 

The quality factor Q of the mechanical oscillator is high $(\omega_{m} \gg \gamma_{m})$. Introducing the slowly varying operators for the linear terms of the fluctuations as $\delta b = {\delta}b e^{-i \omega_{m} t }$, $b_{in} =  b_{in} e^{-i {\omega_{m}} t }$, $\delta c_{1} = {\delta}c_{1} e^{-i  {\Delta_{1}} t }$, ${\delta}c_{2} = {\delta}c_{2} e^{-i {\Delta_{2}} t }$, $c_{1}^{in}=c_{1}^{in} e^{-i\Delta_{1}t}$, $c_{2}^{in}=c_{2}^{in} e^{-i\Delta_{2}t}$, ${\delta}{\sigma^{-}}= {\delta}{\sigma^{-}} e^{-i {\omega_{q}} t }$, ${\sigma^{-}_{in}}={\sigma^{-}_{in}} e^{-i {\omega_{q}} t }$

We thus obtain the linearized quantum Langevin equations for the fluctuations as 

\begin{equation}
\dot{\delta b}= - i {g_{0}} ({c_{2s}}^{\dagger} {\delta}c_{2}  - {c_{1s}}^{\dagger} {\delta}c_{1})- i g {\delta}{\sigma}^{-} - \frac {\gamma_{m}} {2} {\delta}b + \sqrt{\gamma_{m}}  b_{in},
\end{equation}

\begin{equation}
\dot{\delta \sigma}^{-} =  i g {\delta}b {\sigma_{z}} - \frac {k_{d}} {2} {\delta}{\sigma}^{-}(t)+ \sqrt{k_{d}}{ {  \sigma}^{-} }_{in},
\end{equation}

\begin{equation}
\dot{{\delta}c_{1}}=-k {\delta}c_{1}+ i g_{0} c_{1s}{\delta}b + {\epsilon_{L}} {e^{- i x t}} + \sqrt{2k} { c_{1}}^{in},
\end{equation}

\begin{equation}
\dot{{\delta}c_{2}}= -k {\delta}c_{2}- i g_{0} c_{2s}{\delta}b + {\epsilon_{R}}{e^{i{\theta}}}{e^{- i x t}}+ \sqrt{2k} {  c_{2}}^{in}.
\end{equation}

Note that we will be considering $\omega_{m} \gg g_{0} |c_{1s}|$ and  $g_{0} |c_{2s}| $. Here $x= \delta - \omega_{m}$. We take the qubit to be in the ground state i.e, $<\sigma_{z}>=-1$. We now use the ansatz  $<{\delta} s>={\delta} s_{+} {e^{-i{x}t}}+{\delta} s_{-} {e^{i{x}t}}$, with $s= b, c_{1}, c_{2}$ and ${\sigma}^{-}$ . Under steady state condition $<\delta \dot{s}> =0$, we obtain the following expressions,

\begin{equation}
{\delta} b_{+}=\frac{ -i G ( - i x + {\frac{k_{d}}{2}})( n {\epsilon_{R}} {e^{i{\theta}}} - {\epsilon_{L}})}{( - i x + k )  \left [ ( - i x + {\frac{\gamma_{m}}{2}})(- i x + {\frac{k_{d}}{2}}) - g^{2}  {\sigma_{z}} \right ] + (- i x + {\frac{k_{d}}{2}}) G^{2} ( n^{2} + 1)},
\end{equation}

\begin{equation}
{\delta} c_{1+}=\frac{G^{2} \left [ n {\epsilon_{R}} {e^{i{\theta}}} ( - i x + {\frac{k_{d}}{2}}) + n^{2} {\epsilon_{L}} ( - i x + {\frac{k_{d}}{2}}) \right ]+ {\epsilon_{L}} ( - i x + k ) \left [ ( - i x + {\frac{\gamma_{m}}{2}})(- i x + {\frac{k_{d}}{2}}) - g^{2}  {\sigma_{z}} \right ] }{  \left [ ( - i x + {\frac{\gamma_{m}}{2}})(- i x + {\frac{k_{d}}{2}}) - g^{2}  {\sigma_{z}} \right ] ( - i x + k )^{2} + (- i x + {\frac{k_{d}}{2}}) G^{2} ( n^{2} + 1)( - i x + k ) },
\end{equation}

\begin{equation}
{\delta} c_{2+}={\frac{G^{2} \left [ n {\epsilon_{L}} ( - i x + {\frac{k_{d}}{2}} ) +  {\epsilon_{R}}  {e^{i{\theta}}} ( - i x +  {\frac{k_{d}}{2}}) \right ] +  {\epsilon_{R}}  {e^{i{\theta}}} ( - i x + k) \left [ ( - i x + {\frac{\gamma_{m}}{2}})(- i x + {\frac{k_{d}}{2}}) - g^{2}  {\sigma_{z}} \right ] }{  \left [ ( - i x + {\frac{\gamma_{m}}{2}})(- i x + {\frac{k_{d}}{2}}) - g^{2}  {\sigma_{z}} \right ] ( - i x + k )^{2} + (- i x + {\frac{k_{d}}{2}}) G^{2} ( n^{2} + 1)( - i x + k )} }.
\end{equation}

Here $G = g_{0} c_{1s}$ is the effective optomechanical coupling related to coupling power $P_{cL}$. Without loss of generality, we assume $c_{1s}$ and $c_{2s}$ to be real-valued. In addition,  $|{c_{2s} / {c_{1s}}}|= n$, as the photon number ratio of the two cavities.

\subsection{Non-linear case}

Following the procedure adopted in the linear case, we now write down the corresponding equations for the non-linear case. The quantum-Langevin equations are derived as :-

\begin{equation}
\dot{b}= -i {\omega_{m}}b(t) - i {g_{0}}(c_{2}^{\dagger}{c_{2}}-c_{1}^{\dagger}{c_{1}})- {2} i g_{N} b^{\dagger}{\sigma}^{-}(t) - {\frac{\gamma_{m}}{2}}b(t) + \sqrt{\gamma_{m}} b_{in},
\end{equation}

\begin{equation}
\dot{\sigma}^{-}= -i {\omega_{q}}{\sigma}^{-}(t) + i g_{N} b^{2} {\sigma}_{z}(t) - {\frac{k_{d}}{2}}{\sigma}^{-}(t) + \sqrt k_{d}{{\sigma}^{-}_{in}},
\end{equation}

\begin{equation}
\dot{c_{1}}= -[k+ i {\Delta_{c}} -  i {g_{0}}(b^{\dagger}+b)] c_{1} + {\epsilon_{cL}} + {\epsilon_{L}} {e^{-i{\delta}t}}+ \sqrt{2k} {c_{1}}^{in},
\end{equation}

\begin{equation}
\dot{c_{2}}= -[k+ i {\Delta_{c}} +  i {g_{0}}(b^{\dagger}+b)] c_{2} + {\epsilon_{cR}} + {\epsilon_{R}}{e^{i{\theta}}}{e^{-i{\delta}t}}+ \sqrt{2k} {c_{2}}^{in}.
\end{equation}

Note that we have now a non-linear term in the equation for $\sigma^{-}$.

The corresponding steady state values are found from the above equations as 

\begin{equation}
\langle b \rangle = b_{s}=\frac{- i g_{0} ( {|c_{2s}|}^2-{|c_{1s}|}^2)(\frac{k_{d}}{2}+i {\omega_{q}})} { (\frac{\gamma_{m}}{2}+i \omega_{m})(\frac{k_{d}}{2}+i {\omega_{q}})- 2 G_{N}^{2}{\langle \sigma_{z} \rangle}_{s} },
\end{equation}

\begin{equation}
\langle {\sigma^{-}} \rangle= {{\sigma}_{s}}^{-}= \frac{ i g_{N} {b_{s}}^{2} { \langle \sigma_{z} \rangle }_{s}} { (\frac{k_{d}}{2}+ i \omega_{q} ) },
\end{equation}

\begin{equation}
\langle {c_{1}} \rangle =c_{1s}=\frac{\epsilon_{cL}}{k+i{\Delta_{1}}},
\end{equation}

\begin{equation}
\langle {c_{2}} \rangle=c_{2s}=\frac{\epsilon_{cR}}{k+i{\Delta_{2}}},
\end{equation}

where $G_{N}=g_{N} |b_{s}|$ is the effective qubit-mechanical coupling strength. The linearized quantum-Langevin equations for fluctuations are now written as 

\begin{equation}
\dot{\delta b}= - i {g_{0}} ({c_{2s}}^{\dagger} {\delta}c_{2} - {c_{1s}}^{\dagger} {\delta}c_{1})- 2 i g_{N} {b_{s}}^{\dagger} {\delta}{\sigma}^{-} - \frac{\gamma_{m}} {2} {\delta}b + \sqrt{\gamma_{m}}  b_{in},
\end{equation}

\begin{equation}
\dot{ \delta \sigma}^{-}= 2 i g_{N} b_{s} {\sigma_{z}} {\delta}b - {\frac {k_{d}} {2} } {\delta}{\sigma}^{-}(t)+ \sqrt{k_{d}}{{\sigma}^{-}}_{in},
\end{equation}

\begin{equation}
\dot{ \delta c_{1}}=-k {\delta}c_{1}+ i g_{0} c_{1s}{\delta}b + {\epsilon_{L}} {e^{- i x t}} + \sqrt{2k} { c_{1}}^{in},
\end{equation}

\begin{equation}
\dot{\delta c_{2}}= -k {\delta}c_{2}- i g_{0} c_{2s}{\delta}b + {\epsilon_{R}}{e^{i{\theta}}}{e^{- i x t}}+ \sqrt{2k} { c_{2}}^{in}.
\end{equation}

Analogous to eqns.(18)-(20) for the linear case, we now have the corresponding equations for the nonlinear case as 

\begin{equation}
{\delta} b_{+}={\frac{ -i G ( - i x + {\frac{k_{d}}{2}})( n {\epsilon_{R}} {e^{i{\theta}}} - {\epsilon_{L}})}{( - i x + k ) \left [ ( - i x + {\frac{\gamma_{m}}{2}})(- i x + {\frac{k_{d}}{2}}) - {4} {G_{N}}^{2} {\sigma_{z}} \right ] + (- i x + {\frac{k_{d}}{2}}) G^{2} ( n^{2} + 1)}},
\end{equation}

\begin{equation}
{\delta} c_{1+}={\frac{G^{2} \left ( n {\epsilon_{R}} {e^{i{\theta}}} ( - i x + {\frac{k_{d}}{2}}) + n^{2} {\epsilon_{L}} ( - i x + {\frac{k_{d}}{2}}) \right )+ {\epsilon_{L}} ( - i x + k ) \left [ ( - i x + {\frac{\gamma_{m}}{2}})(- i x + {\frac{k_{d}}{2}}) - {4} {G_{N}}^{2} {\sigma_{z}} \right ] }{  \left [ ( - i x + {\frac{\gamma_{m}}{2}})(- i x + {\frac{k_{d}}{2}}) - {4} {G_{N}}^{2} {\sigma_{z}} \right ] ( - i x + k )^{2} + (- i x + {\frac{k_{d}}{2}}) G^{2} ( n^{2} + 1)( - i x + k ) }},
\end{equation}

\begin{equation}
{\delta} c_{2+}={\frac{G^{2} \left ( n {\epsilon_{L}} ( - i x + {\frac{k_{d}}{2}} ) +  {\epsilon_{R}}  {e^{i{\theta}}} ( - i x +  {\frac{k_{d}}{2}}) \right ) +  {\epsilon_{R}}  {e^{i{\theta}}} ( - i x + k) \left [ ( - i x + {\frac{\gamma_{m}}{2}})(- i x + {\frac{k_{d}}{2}}) - {4} {G_{N}}^{2} {\sigma_{z}} \right ] }{ \left [ ( - i x + {\frac{\gamma_{m}}{2}})(- i x + {\frac{k_{d}}{2}}) - {4} {G_{N}}^{2} {\sigma_{z}} \right ] ( - i x + k )^{2} + (- i x + {\frac{k_{d}}{2}}) G^{2} ( n^{2} + 1)( - i x + k ) }}.
\end{equation}

In the next section we will investigate the optical response of the linear and non-linear system.

\section{The Optical Response}

To study the optical response of the system, we utilize the input-output theory \citep{75} and the left hand output field $\epsilon_{out L}$ and the right hand output field $\epsilon_{out R}$ is written as, 
\begin{equation}
{\epsilon_{out L}} + {\epsilon_{L}}{e^{ - i x t }} = 2 k \langle {\delta} c_{1} \rangle,
\end{equation}

\begin{equation}
{\epsilon_{out R}} + {\epsilon_{R}}{e^{ i {\theta}}}{e^{ - i x t }} = 2 k  \langle {\delta} c_{2}\rangle.
\end{equation}

The oscillatory terms can be removed  if we set  ${\epsilon_{out j}} = {\epsilon_{out j+}} {e^{ - i x t }}+{\epsilon_{out j-}} {e^{  i x t }}$,  $( j= L,R)$. Note that the components ${\epsilon_{out L+}}$ and ${\epsilon_{out R+}}$ have same frequency ${\omega_{p}}$ as the probe fields ${\epsilon_{L}}$ and ${\epsilon_{R}}$ while the output components ${\epsilon_{out L-}}$ and ${\epsilon_{out R-}}$ have the frequency $2 {\omega_{c}}- {\omega_{p}}$.

From eqns.(36) and (37) we obtain, 

\begin{equation}
{\epsilon_{out L+}}= 2 k {\delta}c_{1+} - {\epsilon_{L}},
\end{equation}

\begin{equation}
{\epsilon_{out R+}}=2 k {\delta}c_{2+} - {\epsilon_{R}} e^{i\theta}.
\end{equation}

We will now discuss Coherent Perfect Transmission (CPT) and Coherent Perfect Synthesis (CPS) for both the linear and non-linear system.

\subsection{Coherent Perfect Transmission}

We consider here the possibility of acheiving CPT in the parameter regimes where  $|{\frac{{\epsilon_{out}}L+}{{\epsilon_{L}}}}|=0$ and $|{\frac{{\epsilon_{out}}R+}{{\epsilon_{L}}}}|=1$ with ${\epsilon_{L}}{\neq}0$ and $ {\epsilon_{R}}=0$ . These conditions essentially means that we observe the left probe field from the right mirror after it passes through the double cavity system and perfectly transmitted through the right mirror. Note that the right probe field is taken to be absent. 

Taking $\epsilon_{R}=0$ and $n=1$, we get from eqns.(29)-(32) and eqns.(38)-(39), the four points where CPT will occur, in the limit ${\gamma_{m}}, k_{d} \rightarrow {0}$ are 

\begin{equation}
{x_{1} \rightarrow -\frac{\sqrt{2 G^2 + g^2 - k^2 - \sqrt{  4 g^2 k^2 + {(-2 G^2 - g^2 + k^2)}^2}}}{\sqrt{2}}},$$
$${x_{2} \rightarrow \frac{\sqrt{2 G^2 + g^2 - k^2 - \sqrt{4 g^2 k^2 + {(-2 G^2 - g^2 + k^2)}^2}}}{\sqrt{2}}},$$
$${x_{3} \rightarrow -\frac{\sqrt{2 G^2 + g^2 - k^2 + \sqrt{4 g^2 k^2 + {(-2 G^2 - g^2 + k^2)}^2}}}{\sqrt{2}}},$$
$${x_{4} \rightarrow \frac{\sqrt{2 G^2 + g^2 - k^2 + \sqrt{4 g^2 k^2 + {(-2 G^2 - g^2 + k^2)}^2}}}{\sqrt{2}}}.
\end{equation}

We first consider the linear case.

\begin{figure}[ht]
\hspace{-0.0cm}
\begin{tabular}{cc}
\includegraphics [scale=0.60] {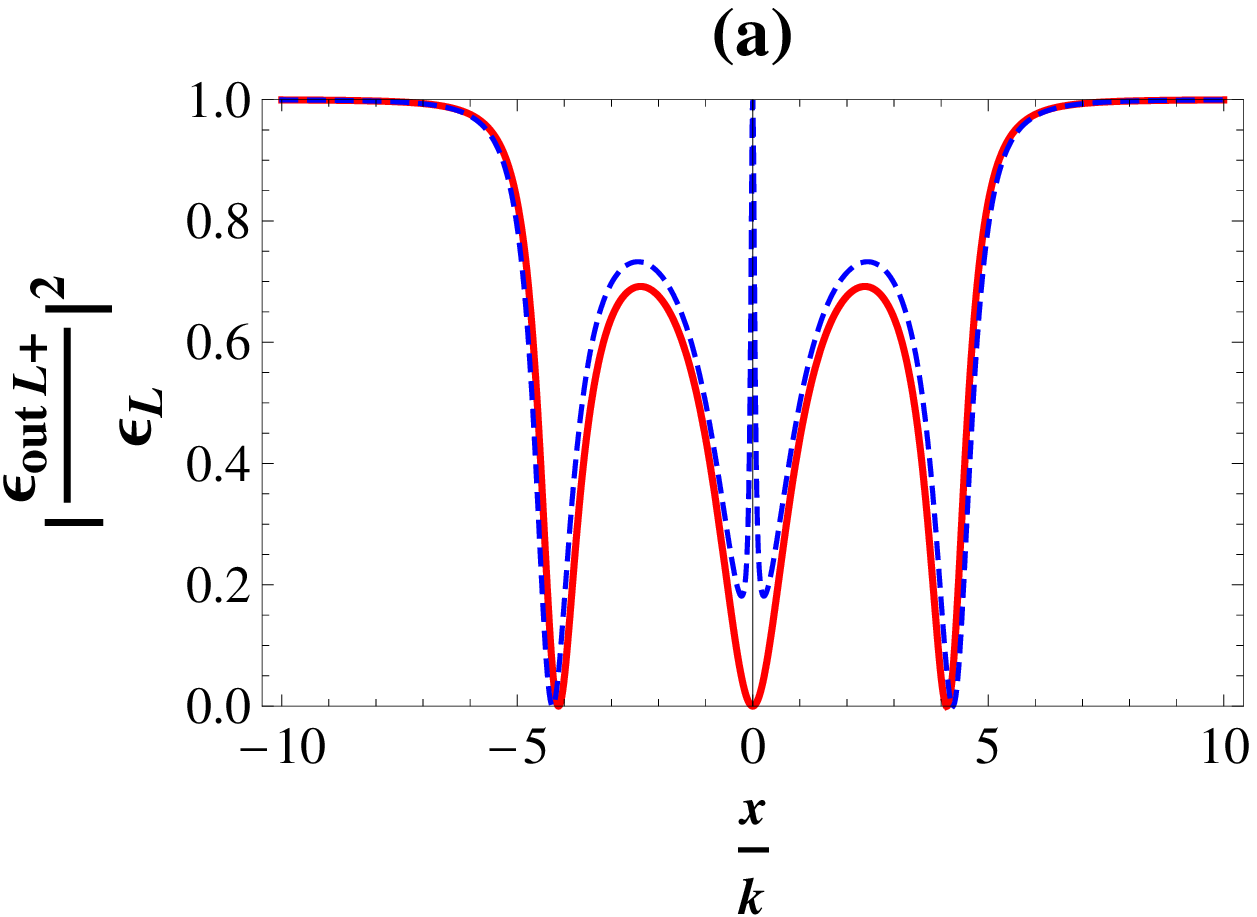} \includegraphics [scale=0.60] {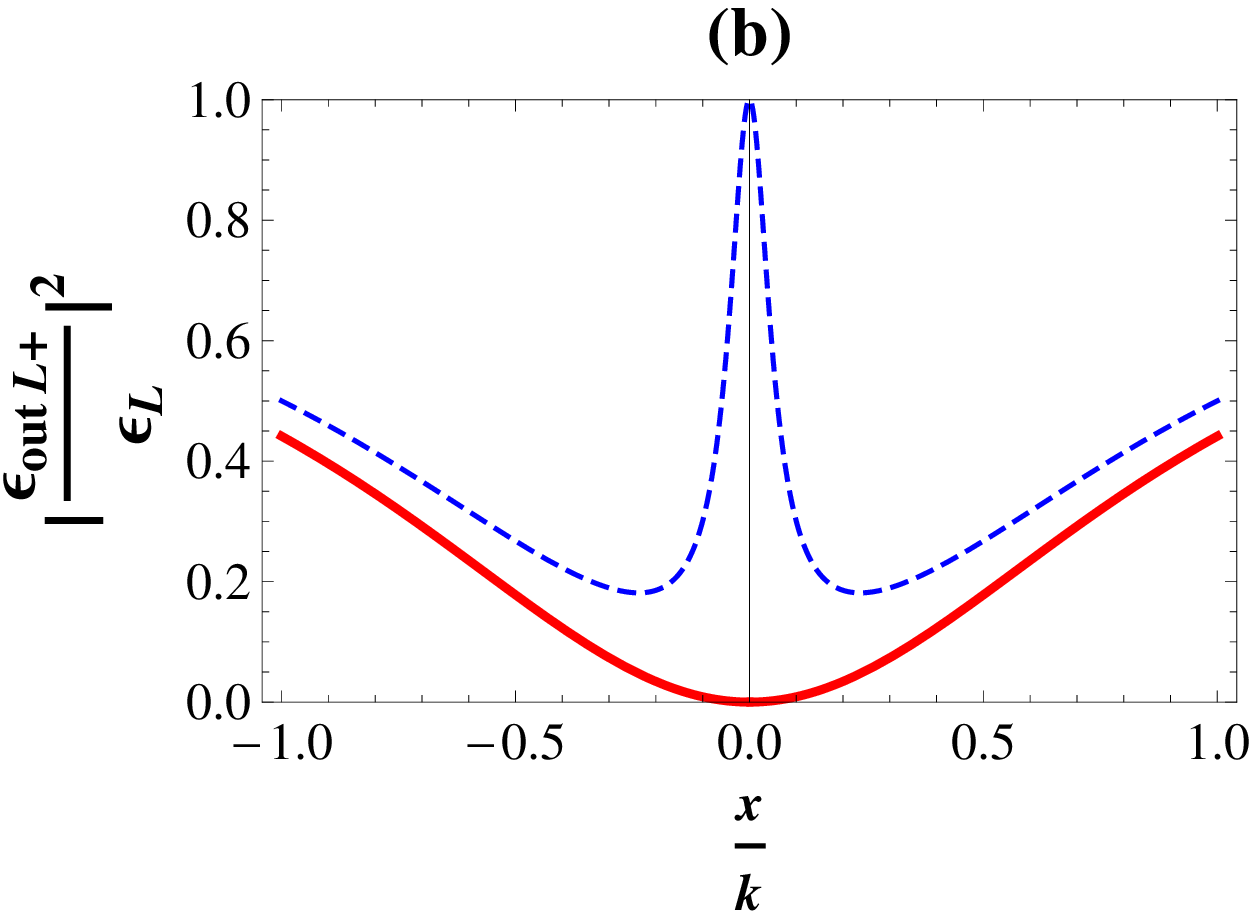}\\
\end{tabular}
\caption{(Color online) The left output field  $|{\frac{{\epsilon_{out}}L+}{{\epsilon_{L}}}}|^{2}$ as a function of normalized probe detuning $x/k$. (a): [$G={{3}{k}}$, $g={0}$ (Solid red-line)]; [$G={{3}{k}}$, $g={k}$ (Dashed blue-line)]; (b): Same plot as in (a) near $x/k=0$ shown for clarity.}
\label{Fig1(a,b)}

\end{figure}

\begin{figure}[ht]
\hspace{-0.0cm}
\begin{tabular}{cc}
\includegraphics [scale=0.60] {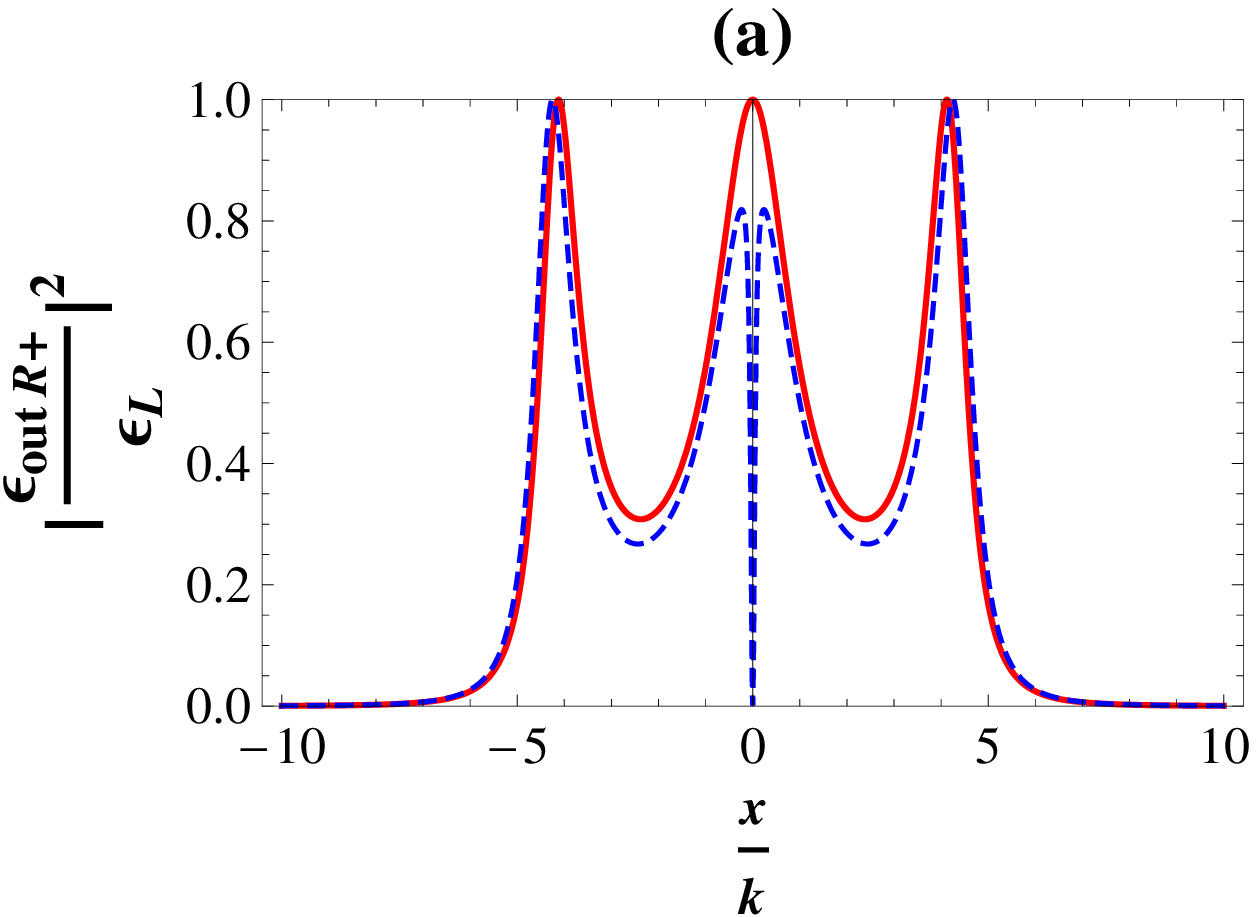} \includegraphics [scale=0.60] {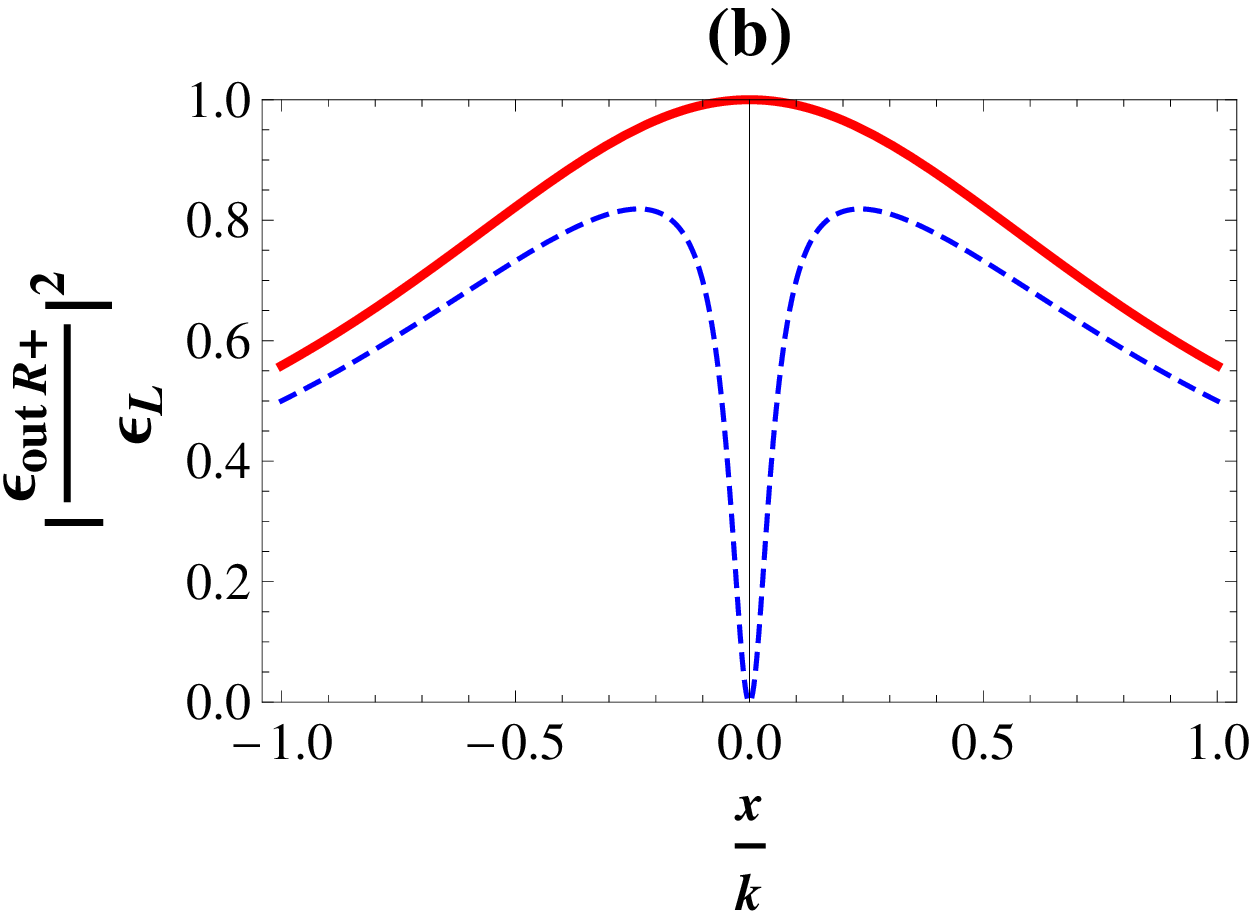}\\
\end{tabular}
\caption{(Color online) The right output field  $|{\frac{{\epsilon_{out}}R+}{{\epsilon_{L}}}}|^{2}$ as a function of normalized probe detuning $x/k$. (a): [$G={{3}{k}}$, $g={0}$ (Solid red-line)]; [$G={{3}{k}}$, $g={k}$ (Dashed blue-line)]; (b): Same plot as in (a) near $x/k=0$.}
\label{Fig2(a,b)}

\end{figure}

In fig.2 and fig.3, we plot the normalized output probe field energy $|{\frac{{\epsilon_{out}}L+}{{\epsilon_{L}}}}|^{2}$ and $|{\frac{{\epsilon_{out}}R+}{{\epsilon_{L}}}}|^{2}$ respectively, as a function of dimensionless input probe detuning $x/k$ for $G={{3}{k}}$; $g=0$ (Solid red-line), and $G={{3}{k}}$; $g={k}$ (Dashed blue-line). Fig. 2(b) and fig. 3(b) shows the same plots near $x/k=0$ for clarity. For $G={{3}{k}}$ and  $g=0$, we get three points where CPT is observed i.e, $x/k=0$ and  
$x_{\pm}={\pm}{4.05}{k}$. On the other hand for $G={{3}{k}}$ and $g={k}$, four distinct transmission points are noticed both from the plots as well as from eqn.(40). In the absence of two-level-mechanical mode coupling ($g=0$), CPT is observed at three points but on the other hand when g is finite i.e, $g=k$, the two points near $x/k$ do not demonstrate CPT. This perhaps indicates that some energy from the optical mode is taken away by the mechanical mode via optomechanical coupling and transfered to the two-level system.
The two points near $x_{\pm}={\pm}{4.05}{k}$ shows perfect transmission (CPT). 

\begin{figure}[ht]
\hspace{-0.0cm}
\begin{tabular}{cc}
\includegraphics [scale=0.60] {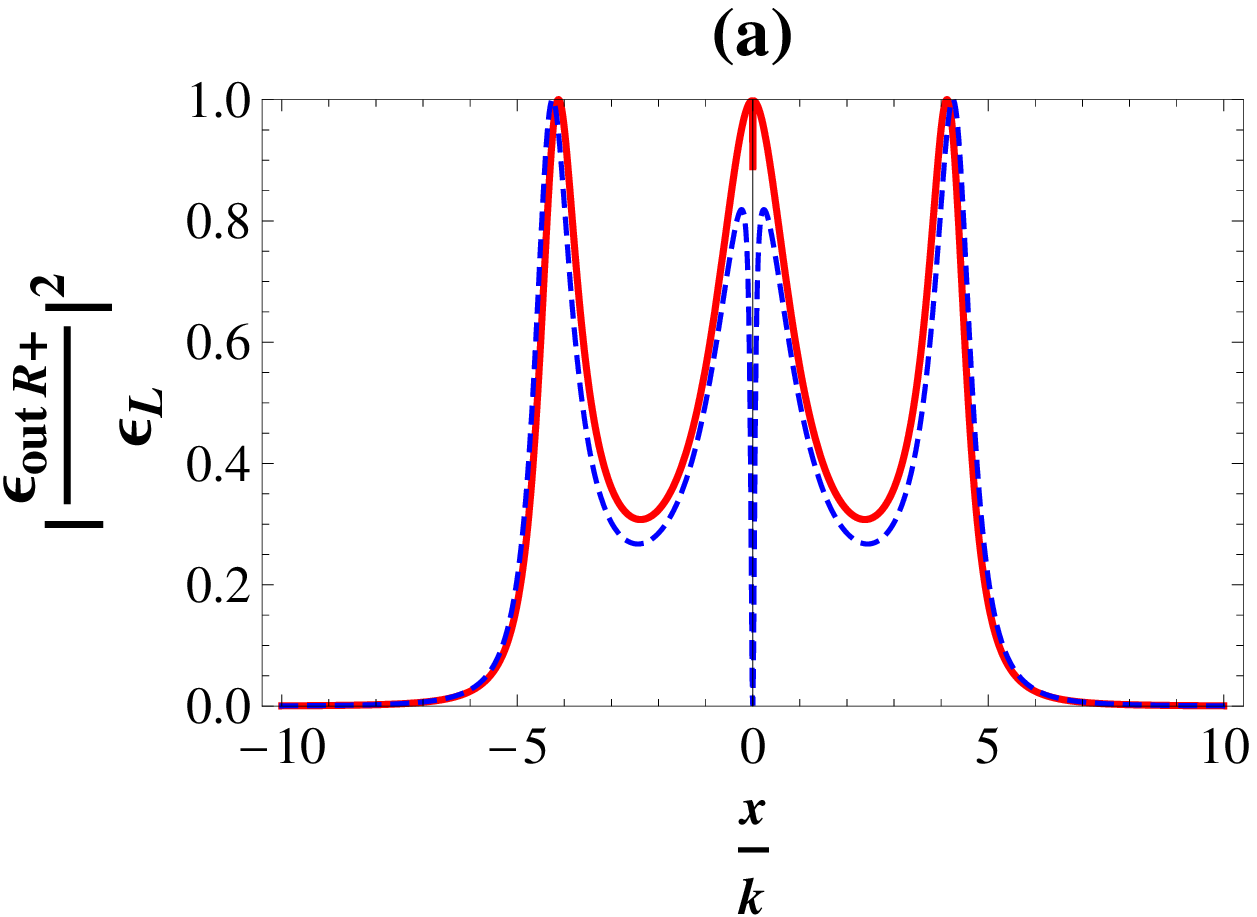} \includegraphics [scale=0.60] {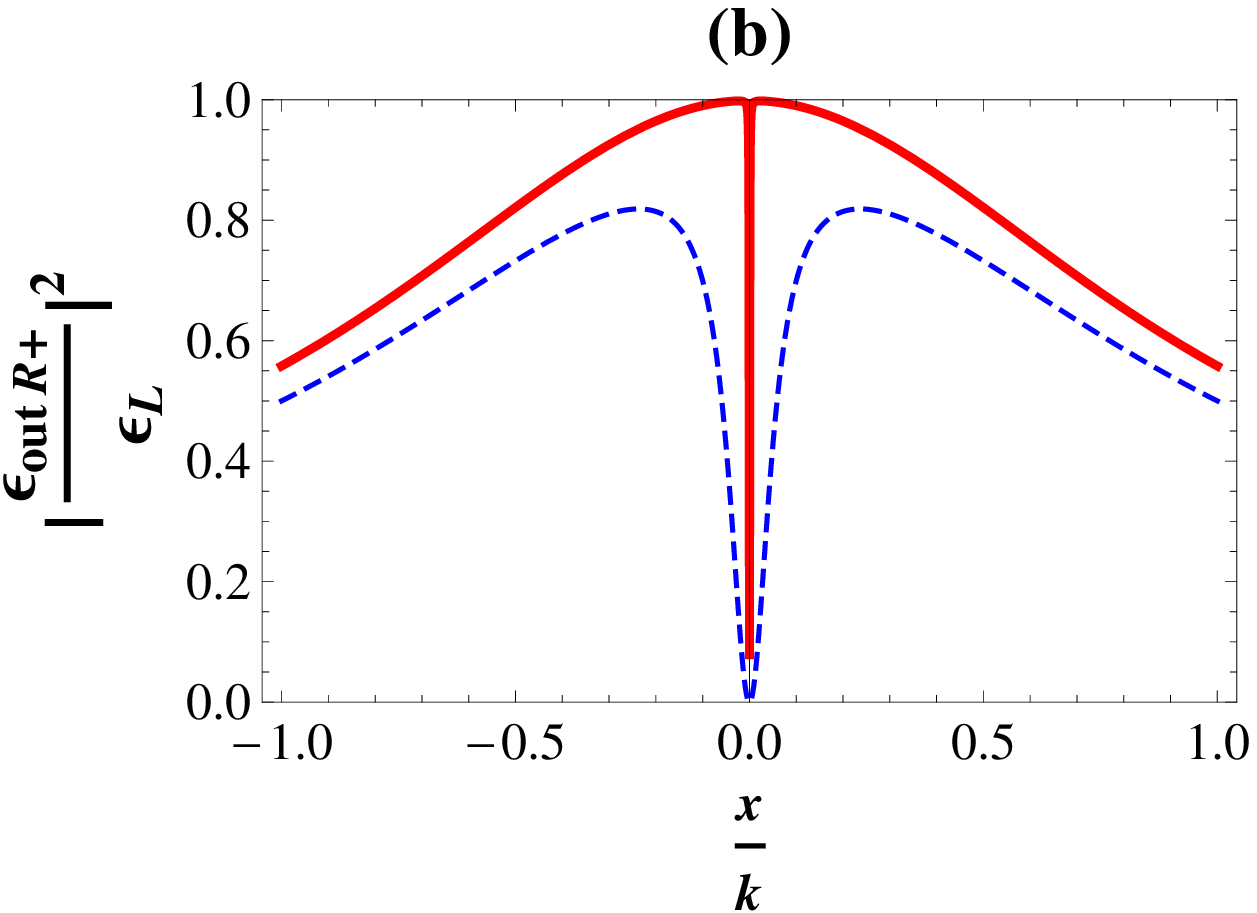}\\
\end{tabular}
\caption{(Color online) The right output field  $|{\frac{{\epsilon_{out}}R+}{{\epsilon_{L}}}}|^{2}$ as a function of normalized probe detuning $x/k$. (a) [$G={{3}{k}}$, $g={0.1}k$ (Solid red-line)]; [$G={{3}{k}}$, $g={k}$ (Dashed blue-line)]; (b) Same plot as in (a) near $x/k=0$ for clarity.}
\label{Fig3(a,b)}
\end{figure}

Fig.4 shows a plot of $|{\frac{{\epsilon_{out}}R+}{{\epsilon_{L}}}}|^{2}$ for $g=0.1 k$ (solid red-line) and $g=k$ (dashed blue-line). Even at $g=0.1k$, only three CPT points are visible. Actually four transmission points starts appearing when $g \geq 0.4k$.

We now consider the non-linear case corresponding to a superconducting charge qubit  attached to the membrane in the middle. Proceeding in a manner similar to that for the linear case, the four transmission point appear at,

\begin{equation}
{x_{1} \rightarrow -\frac{\sqrt{2 G^2 + 4 G_{N}^2 - k^2 - \sqrt{  16 G_{N}^2 k^2 + {(-2 G^2 - 4 G_{N}^2 + k^2)}^2}}}{\sqrt{2}}},$$
$${x_{2} \rightarrow \frac{\sqrt{2 G^2 + 4 G_{N}^2 - k^2 - \sqrt{16 G_{N}^2 k^2 + {(-2 G^2 - 4 G_{N}^2 + k^2)}^2}}}{\sqrt{2}}},$$ 
$${x_{3} \rightarrow -\sqrt{ G^2 + 2 G_{N}^2 - \frac{k^2}{2} + \frac{\sqrt   {16 G_{N}^2 k^2 + {(-2 G^2 -4 G_{N}^2 + k^2)}^2}  }{2}}},$$
$${x_{4} \rightarrow \sqrt{ G^2 +2 G_{N}^2 -\frac{k^2}{2} +\frac{\sqrt{16 G_{N}^2 k^2 + {(-2 G^2 - 4 G_{N}^2 + k^2)}^2}  }  {2} }},
\end{equation}

\begin{figure}[ht]
\hspace{-0.0cm}
\begin{tabular}{cc}
\includegraphics [scale=0.60] {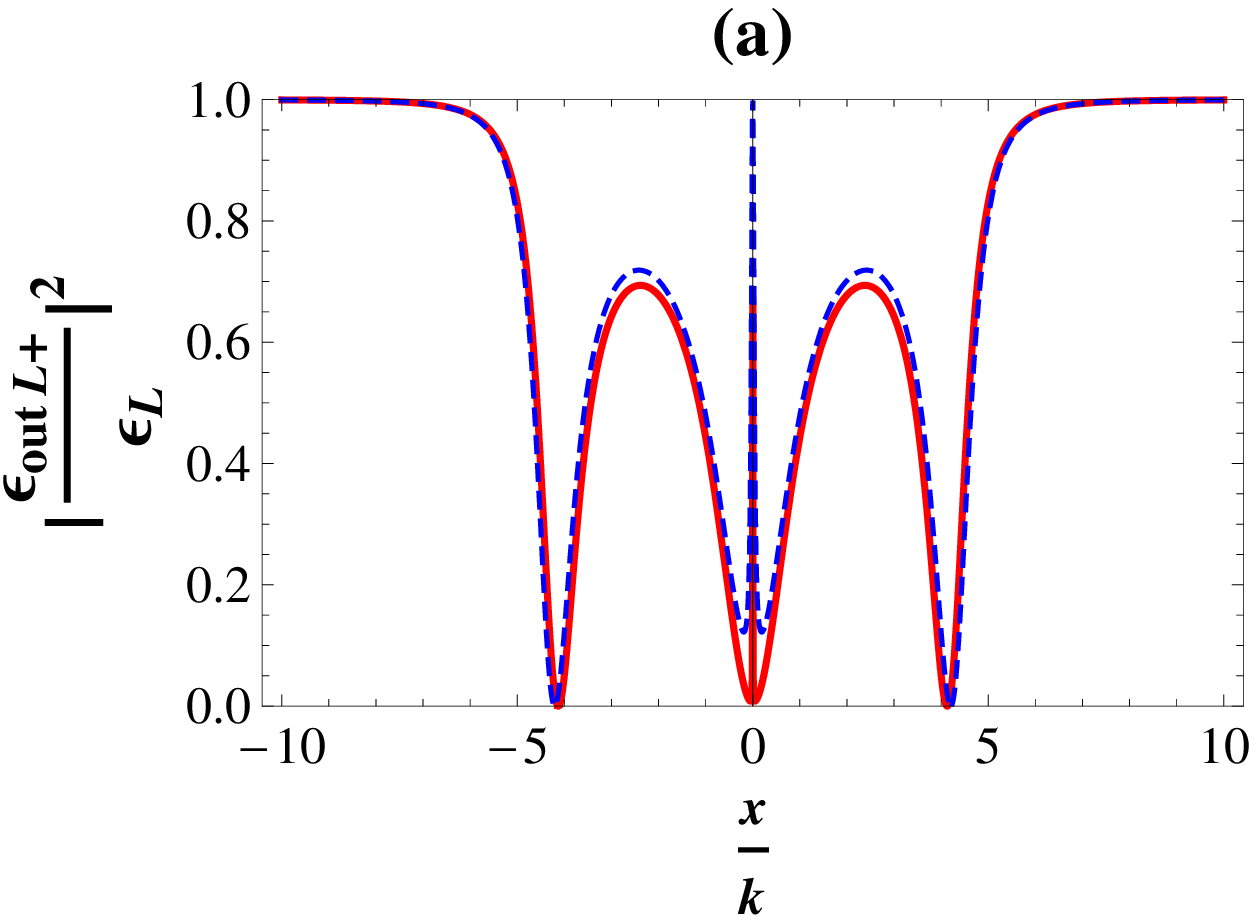} \includegraphics [scale=0.60] {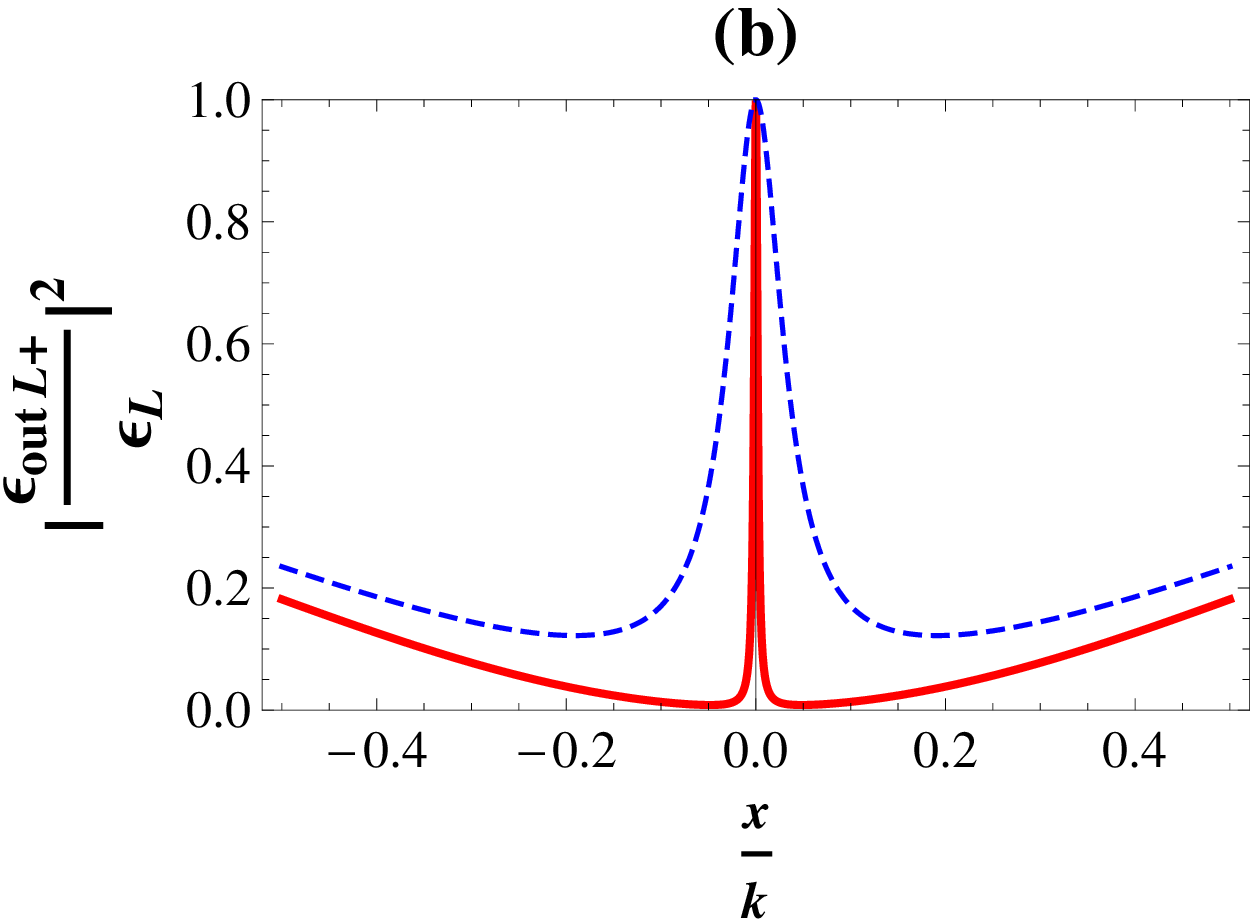}\\
\end{tabular}
\caption{(Color online) The left output field $|{\frac{{\epsilon_{out}}L+}{{\epsilon_{L}}}}|^{2}$ as a function of normalized probe detuning $x/k$. (a) [$G={{3}{k}}$, $G_{N}={{0.1}{k}}$ (Solid red-line)], [ $G={{3}{k}}$, $G_{N}={{0.4}{k}}$ (Dashed blue-line)], (b) Same plot as in (a) near $x/k=0$.}
\label{Fig 4(a,b)}

\end{figure}

\begin{figure}[ht]
\hspace{-0.0cm}
\begin{tabular}{cc}
\includegraphics [scale=0.60] {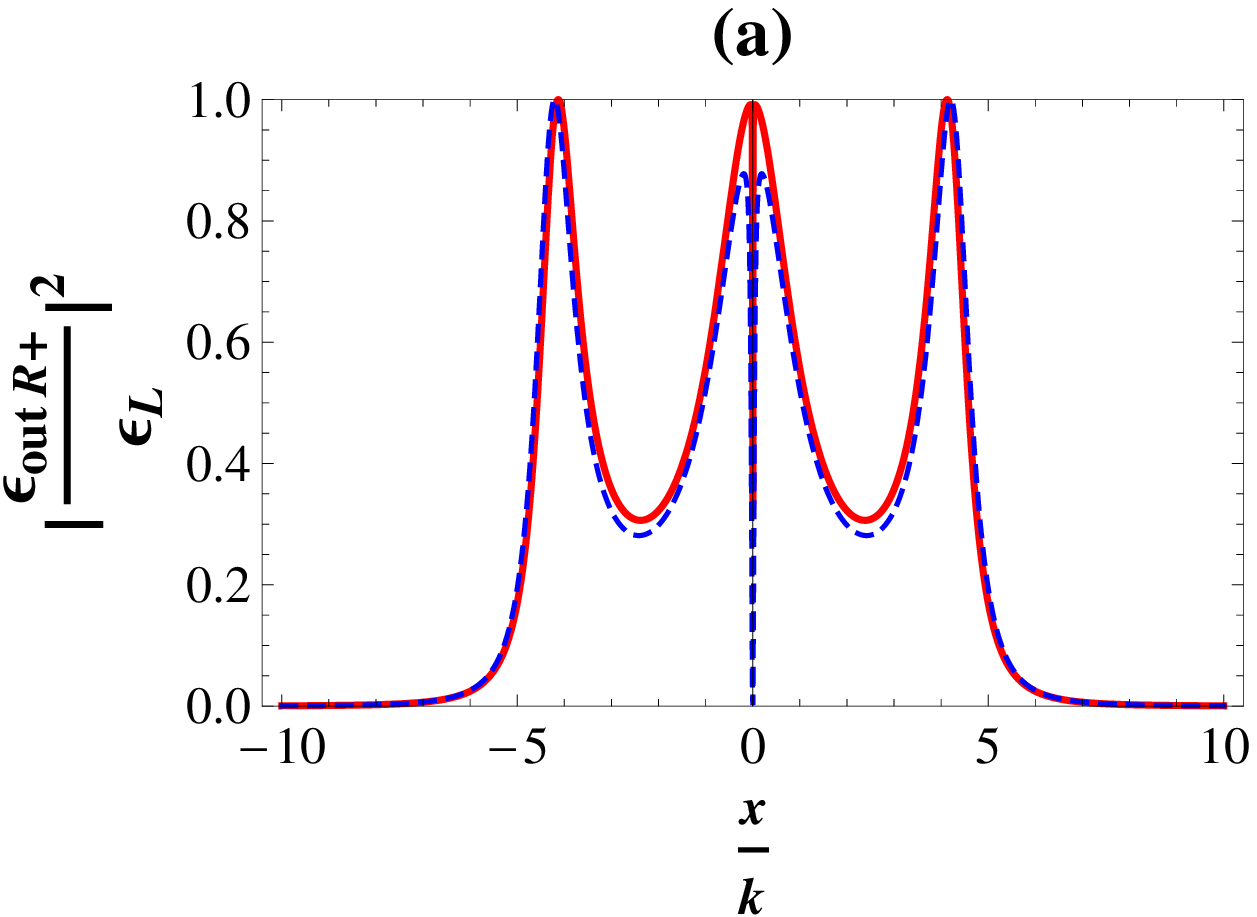} \includegraphics [scale=0.60] {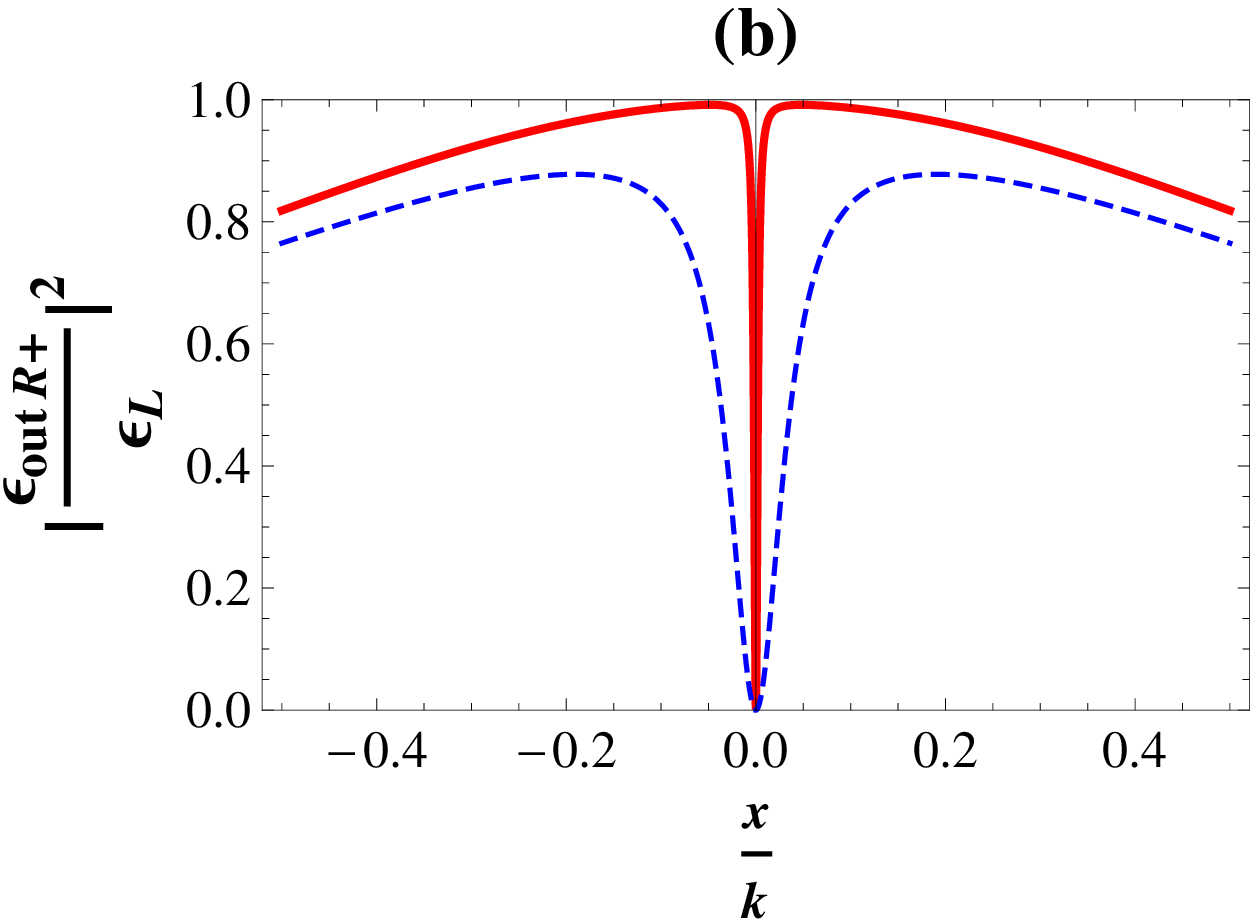}\\
\end{tabular}
\caption{(Color online)  The right output field  $|{\frac{{\epsilon_{out}}R+}{{\epsilon_{L}}}}|^{2}$ as a function of normalized probe detuning $x/k$. (a) [$G={{3}{k}}$, $G_{N}={{0.1}{k}}$ (Solid red-line)], [ $G={{3}{k}}$, $G_{N}={{0.4}{k}}$ (Dashed blue-line)], (b) Same plot as in (a) near $x/k=0$.}
\label{Fig 5(a,b)}

\end{figure}

Fig. 5 and fig. 6 shows the plots of $|{\frac{{\epsilon_{out}}L+}{{\epsilon_{L}}}}|^{2}$ and $|{\frac{{\epsilon_{out}}R+}{{\epsilon_{L}}}}|^{2}$ respectively, as a function of input probe detuning ${\frac{x}{k}}$ for $G={{3}{k}}$; $G_{N}=0.1 k$ (Red-line), and $G={{3k}}$; $G_{N}=0.4{k}$ (Blue Dashed-line). Now comparing fig.3 and fig.5, we note that CPT at four points is observed at $G_{N}=0.1 k$ while at $G_{N}=0.4{k}$, complete transmission is only observed at two points near $x_{\pm}={\pm}{4}{k}$. The two points near $x=0$ show near perfect transmission ($|{\frac{{\epsilon_{out}}R+}{{\epsilon_{L}}}}|^{2} \approx 0.85 $) for $G_{N}=0.
4k$ which is slightly higher than $g=k$ case.

For all the above cases, exactly at $x=0$, CPT is observed only when $g(G_{N})=0$. Thus in the presence of the two-level system, we can design an ``All Optical Switch'' functioning around $x=0$.
From the above analysis, we can conclude that the non-linear system is comparatively more suitable to generate four CPT points. Around $x=0$ points both for linear as well as non-linear case, we observe two transmission points. As one tunes $x$ across the $x=0$ point, we can switch between zero transmission to large transmission ( CPT in case $g$ or $G_{N}$ is very low ). Thus these systems have the potential to be used as ``all optical switch''. Moreover, the width of the transmission around $x=0$ is very small for small $g (G_{N})$ and it widens as we increase $g (G_{N})$.This indicates that at small values of $g$ or $G_{N}$, the functioning of the optical switch is more sensitive i.e, a small variation of $x$ around $x=0$ causes a sharp change in the transmission. 

\subsection{Coherent Perfect Synthesis}

In this sub-section, we consider the possibility of achieving CPS under the conditions $|{\frac{{\epsilon_{out}}L+}{{\epsilon_{L}}}}|^{2}=0$ and $|{\frac{{\epsilon_{out}}R+}{{\epsilon_{L}}}}|^{2}=2$ or $|{\frac{{\epsilon_{out}}L+}{{\epsilon_{L}}}}|^{2}=2$ and $|{\frac{{\epsilon_{out}}R+}{{\epsilon_{L}}}}|^{2}=0$ with ${\epsilon_{L}}=\epsilon_{R} {\neq} 0 $. In order to avoid energy loss via fast mechanical decay, we consider a high-Q quantum mechanical mode by taking ${\gamma_{m}} \rightarrow {0}$. In addition we also assume that energy is not loss due to decay and decoherence of the two level system by taking $k_{d} \rightarrow 0$. A plot of $|{\frac{{\epsilon_{out}}L+}{{\epsilon_{L}}}}|^{2}$ and $|{\frac{{\epsilon_{out}}R+}{{\epsilon_{R}}}}|^{2}$ versus normalized detuning $x/k$ for the linear case is shown in fig.7(a) and fig.7(b) respectively. 

\begin{figure}[ht]
\hspace{-0.0cm}
\begin{tabular}{cc}
\includegraphics [scale=0.60] {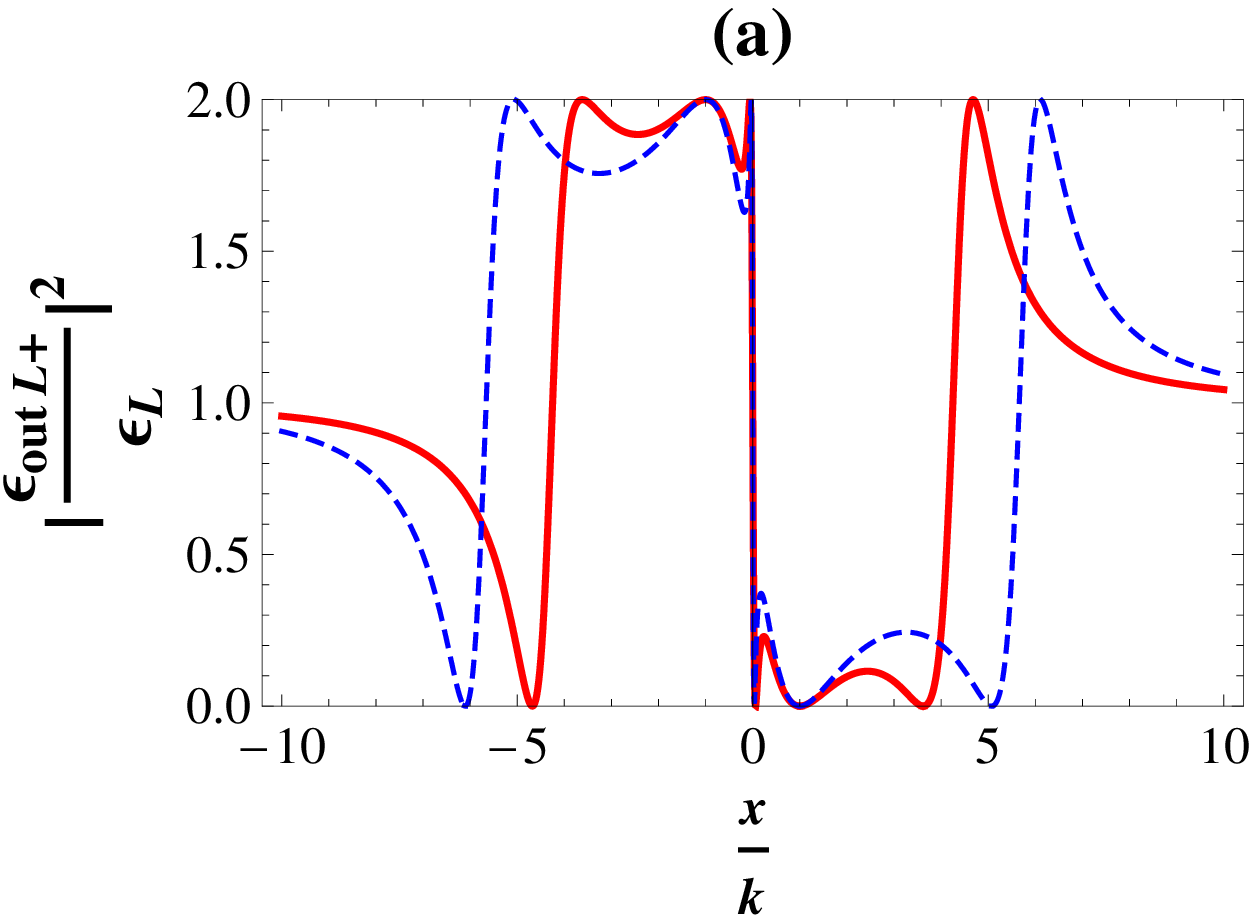} \includegraphics [scale=0.60] {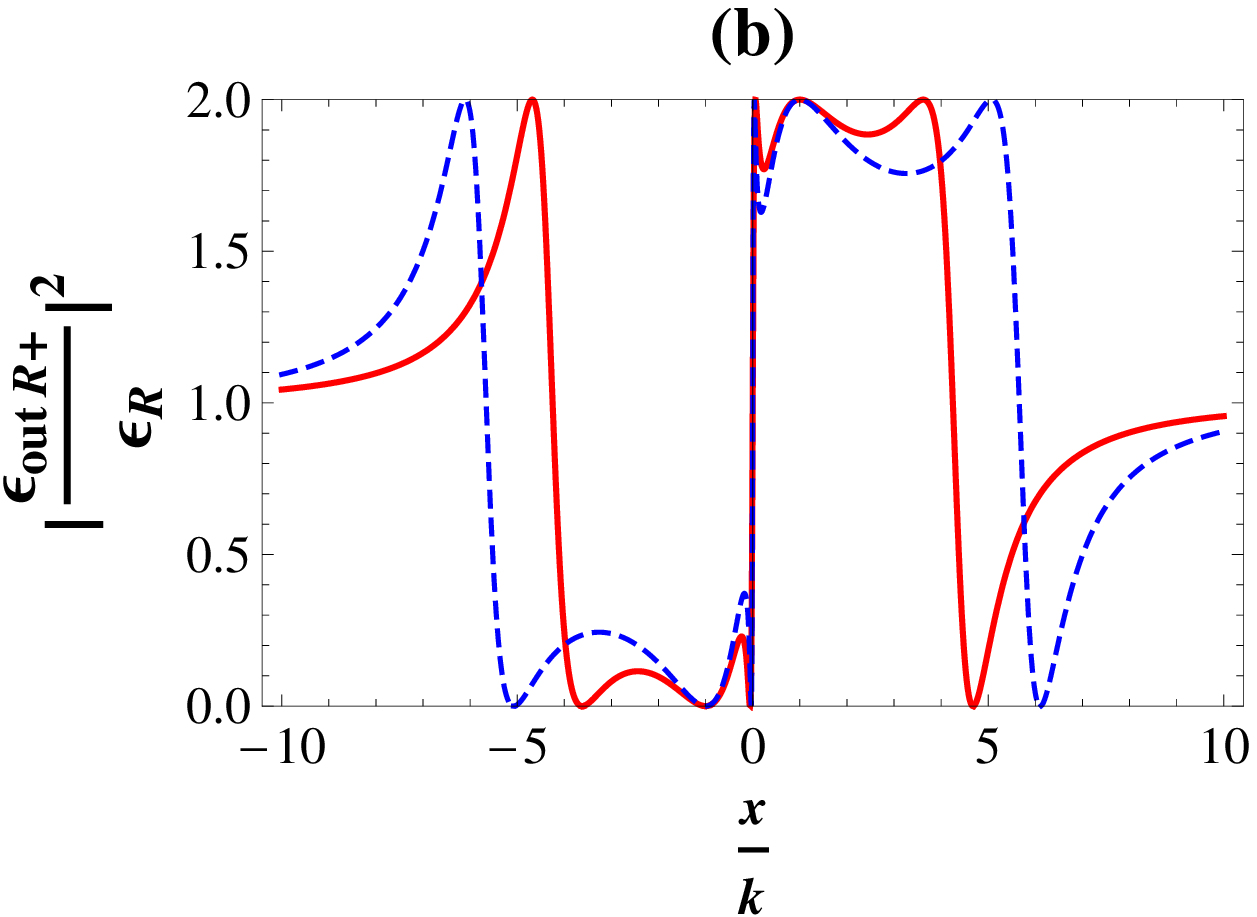}\\
\end{tabular}

\caption{(Color online) The normalized output strength for linear case for (a) $|{\frac{{\epsilon_{out}}L+}{{\epsilon_{L}}}}|^{2}$ and  (b) $|{\frac{{\epsilon_{out}}R+}{{\epsilon_{R}}}}|^{2}$ as a function of normalized probe detuning $x/k$.  The parameters used are: $G={{3}{k}}$ and $g={k}$ (Solid red-line), $G={{4}{k}}$ and $g={k}$ (Dashed blue-line)]}.
\label{Fig6 (a,b)}

\end{figure}

Four perfect synthesis channels are produced for $g=k$, $G=3k$ (solid red-line) as well as for $G=4k$ (dashed blue-line). It is clear that points where $|{\frac{{\epsilon_{out}}L+}{{\epsilon_{L}}}}|^{2}=0$, we have $|{\frac{{\epsilon_{out}}R+}{{\epsilon_{R}}}}|^{2}=2$ and vice-versa. 

Plots for the non-linear case is depicted in fig.8(a) and 8(b) respectively. A similar observation is made as in the linear case (fig.7). Four perfect synthesis points are visible here also. 

\begin{figure}[ht]
\hspace{-0.0cm}
\begin{tabular}{cc}
\includegraphics [scale=0.60] {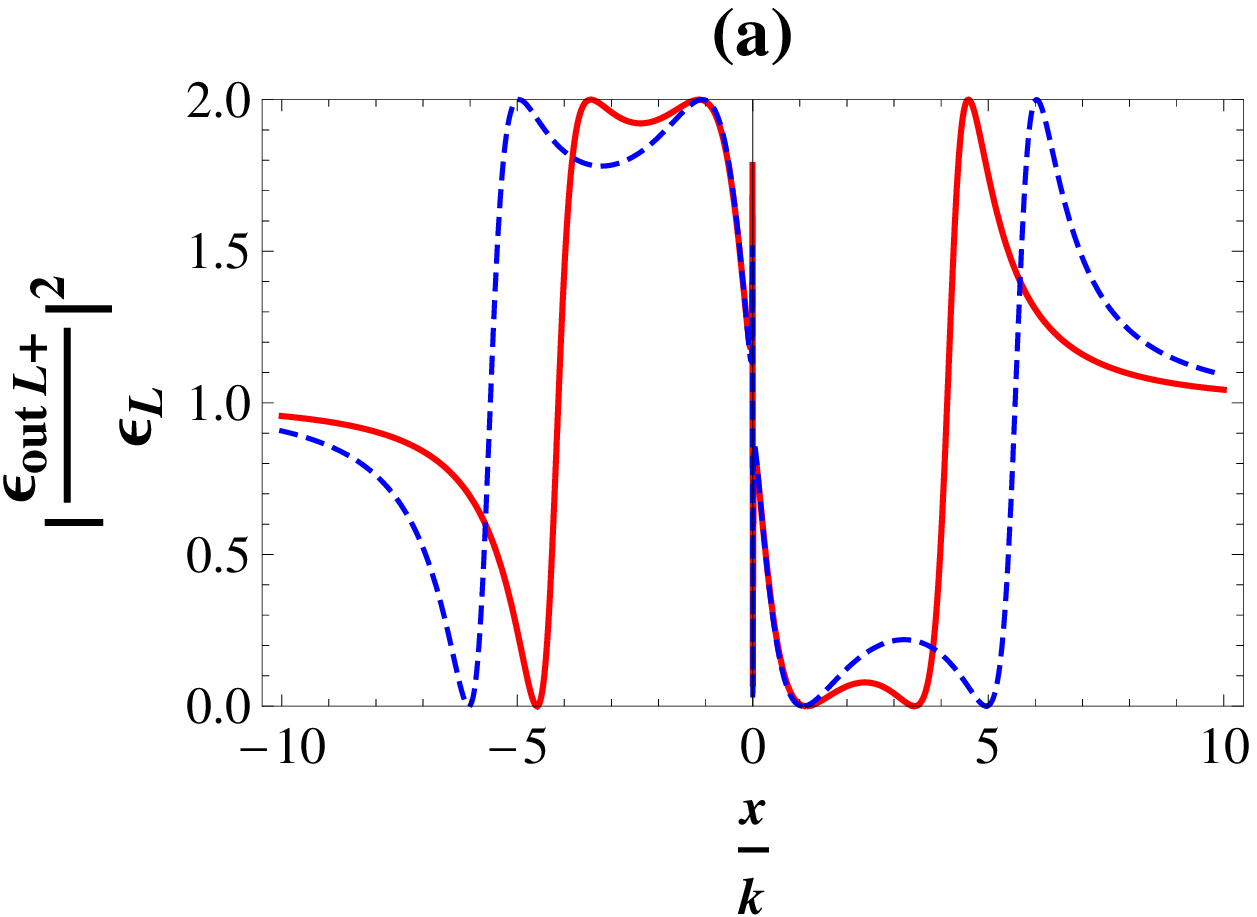} \includegraphics [scale=0.60] {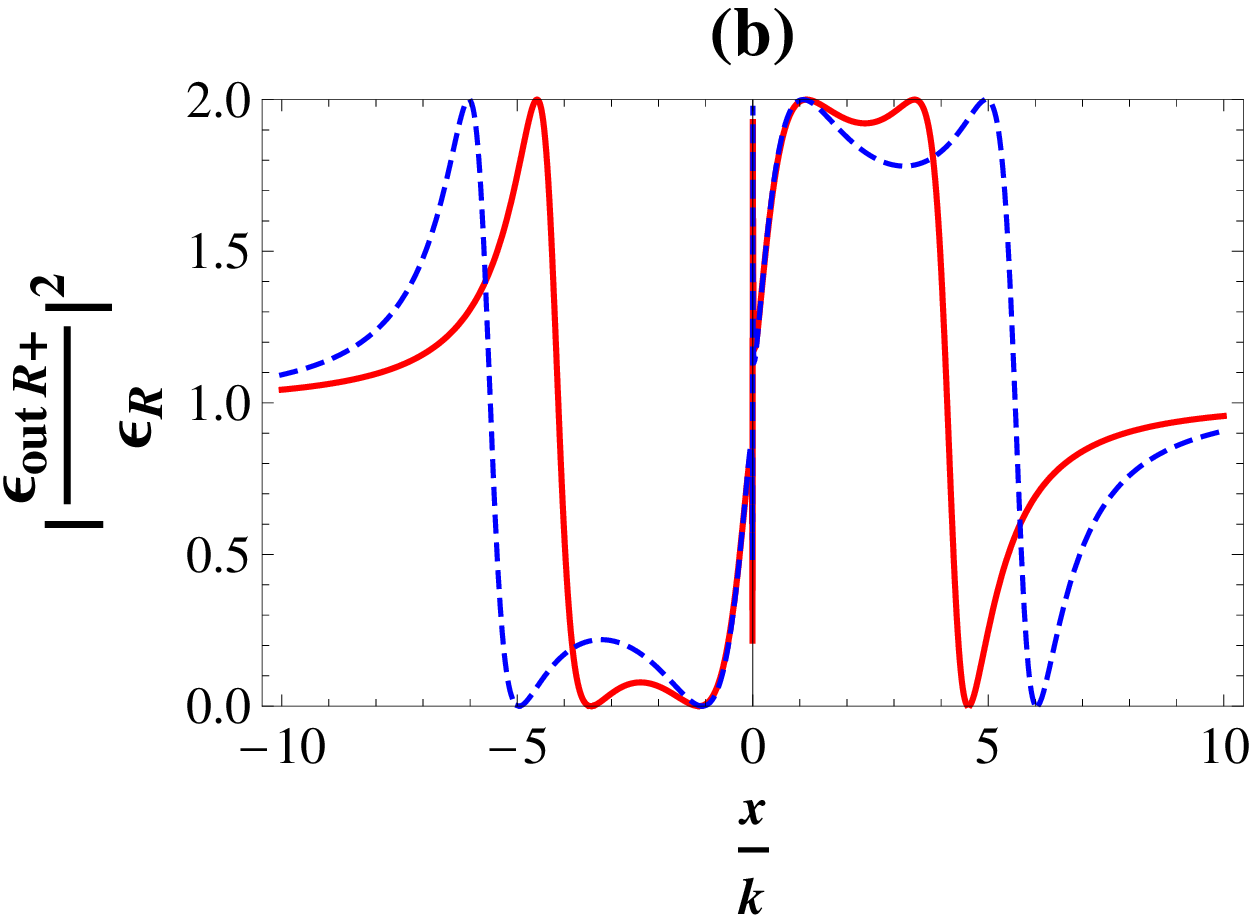}\\
\end{tabular}
\caption{(Color online) The normalized output strength for non-linear case for (a) $|{\frac{{\epsilon_{out}}L+}{{\epsilon_{L}}}}|^{2}$ and (b) $|{\frac{{\epsilon_{out}}R+}{{\epsilon_{R}}}}|^{2}$ as a function of normalized probe detuning $x/k$. The parameters used are:  $G={{3}{k}}$ and $G_{N}={{0.1}{k}}$ (Solid red-line) ,  $G={{4}{k}}$ and  $G_{N}={{0.1}{k}}$ (dashed blue-line).}
\label{Fig7 (a,b)}

\end{figure}

Fig.9 illustrates both the linear and nonlinear case with the reduced range of $x/k$. This helps us to focus around the $x=0$ point. Interestingly we notice that the variation of the output probe energy for the nonlinear case is extremely rapid around $x=0$ as compared to the linear case. This again as before points to the fact that the nonlinear system can be used to design a comparatively more sensitive optical switch. 

Thus we see that in CPS we can have a coherent control over the perfect transmission and perfect reflection of the left and right probe fields. These observations are a result of constructive or destructive interference between $\epsilon_{L}$ and $\epsilon_{R}$ at the two cavity mirrors. This interference is seen to be influenced by the presence of the two-level system coupled to the middle movable membrane and that we can control the transmission by the adjusting the two-level parameters which is seen to emerge as a new handle.

\begin{figure}[ht]
\hspace{-0.0cm}
\begin{tabular}{cc}
\includegraphics [scale=0.60] {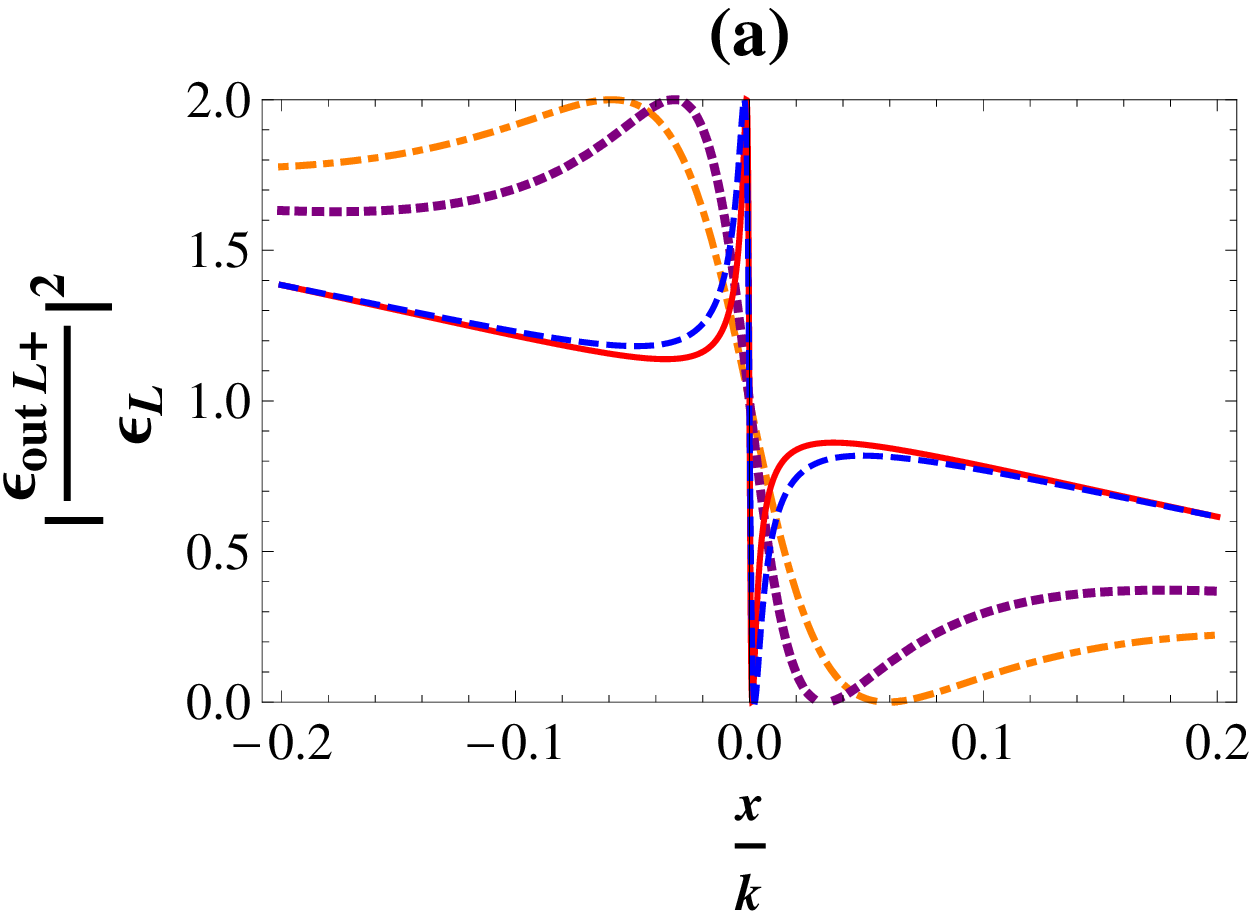} \includegraphics [scale=0.60] {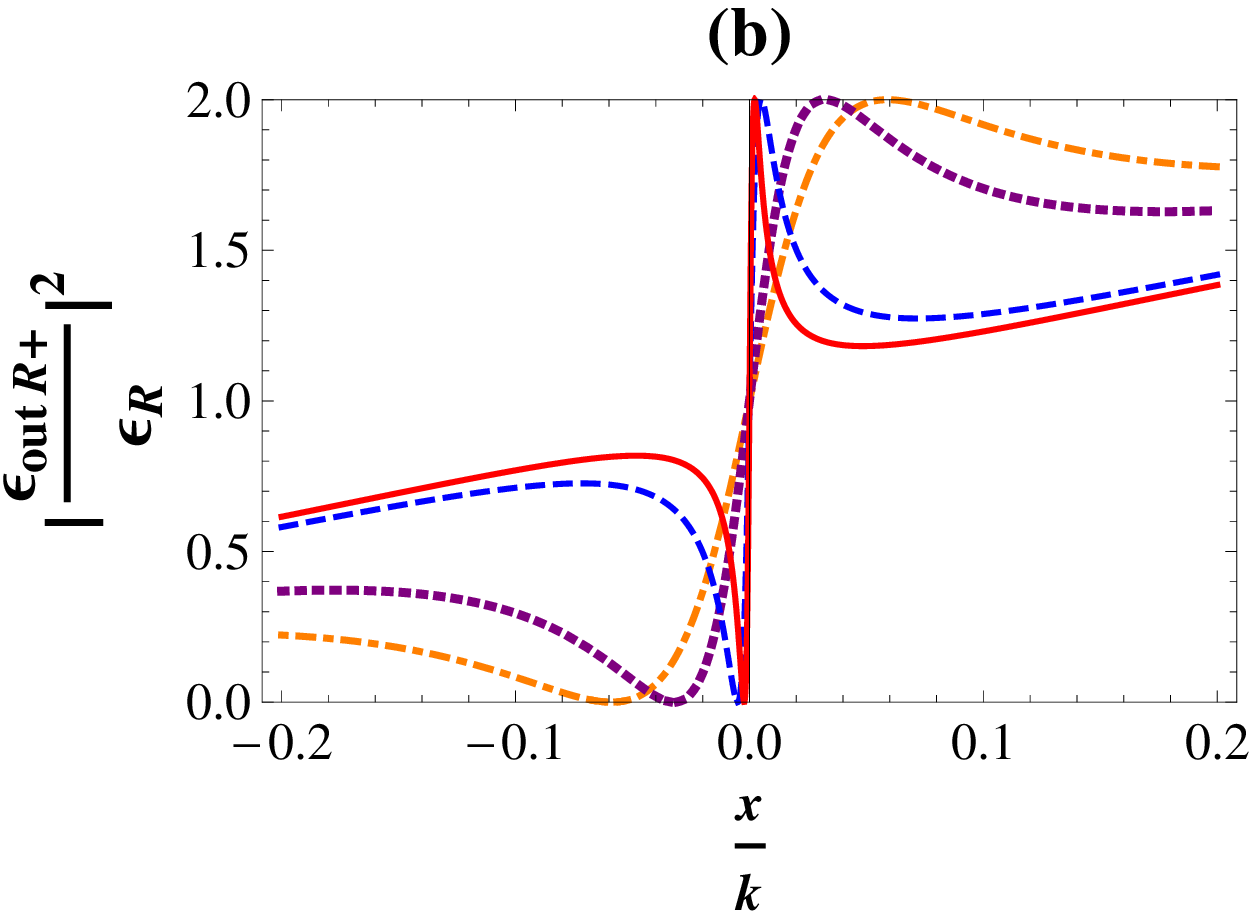}\\
\end{tabular}

\caption{(Color online)The normalized output strength for both linear and non-linear case for (a) $|{\frac{{\epsilon_{out}}L+}{{\epsilon_{L}}}}|^{2}$ and  (b) $|{\frac{{\epsilon_{out}}R+}{{\epsilon_{R}}}}|^{2}$ as a function of normalized probe detuning $x/k$ at near $x/k=0$ for the following parameters: [$G={{3}{k}}$, $g={k}$ (Orange dot-dashed-line)], [$G={{4}{k}}$, $g={k}$ (Purple dotted-line)], [$G={{3}{k}}$, $G_{N}={{0.1}{k}}$ (Blue dashed-line) ], [$G={{4}{k}}$, $G_{N}={{0.1}{k}}$ (Red solid-line)] ;}
\label{Fig8 (a,b)}

\end{figure}

\section{Optomechanically Induced Absorption}

In optomechanically induced transparency (OMIT), probe excitations are transferred to mechanical oscillations and again converted back to probe field. A perfect destructive interference can be set up between the intracavity probe field and the fluctuations that returns to the cavity from the mechanical oscillator. As a result the probe field can not exist in the cavity, and the cavity then becomes transparent. A system can also be designed such that a constructive interference take place that leads to optomechanically induced absorption (OMIA) \citep{76,77}. In this section, we analyze the existence of OMIA in terms of the left-hand or right-hand output probe fields. The absorptive and dispersive behaviour of the system is contained in the real and imaginary part of the transmission $\epsilon_{T}$. Defining $\epsilon_{T}={\dfrac{2k{\delta}c_{1+}}{{\epsilon_{L}}}},$ we obtain the following expressions for linear coupling 

\begin{figure}[ht]
\hspace{-0.0cm}

\begin{tabular}{cc}
\includegraphics [scale=0.60] {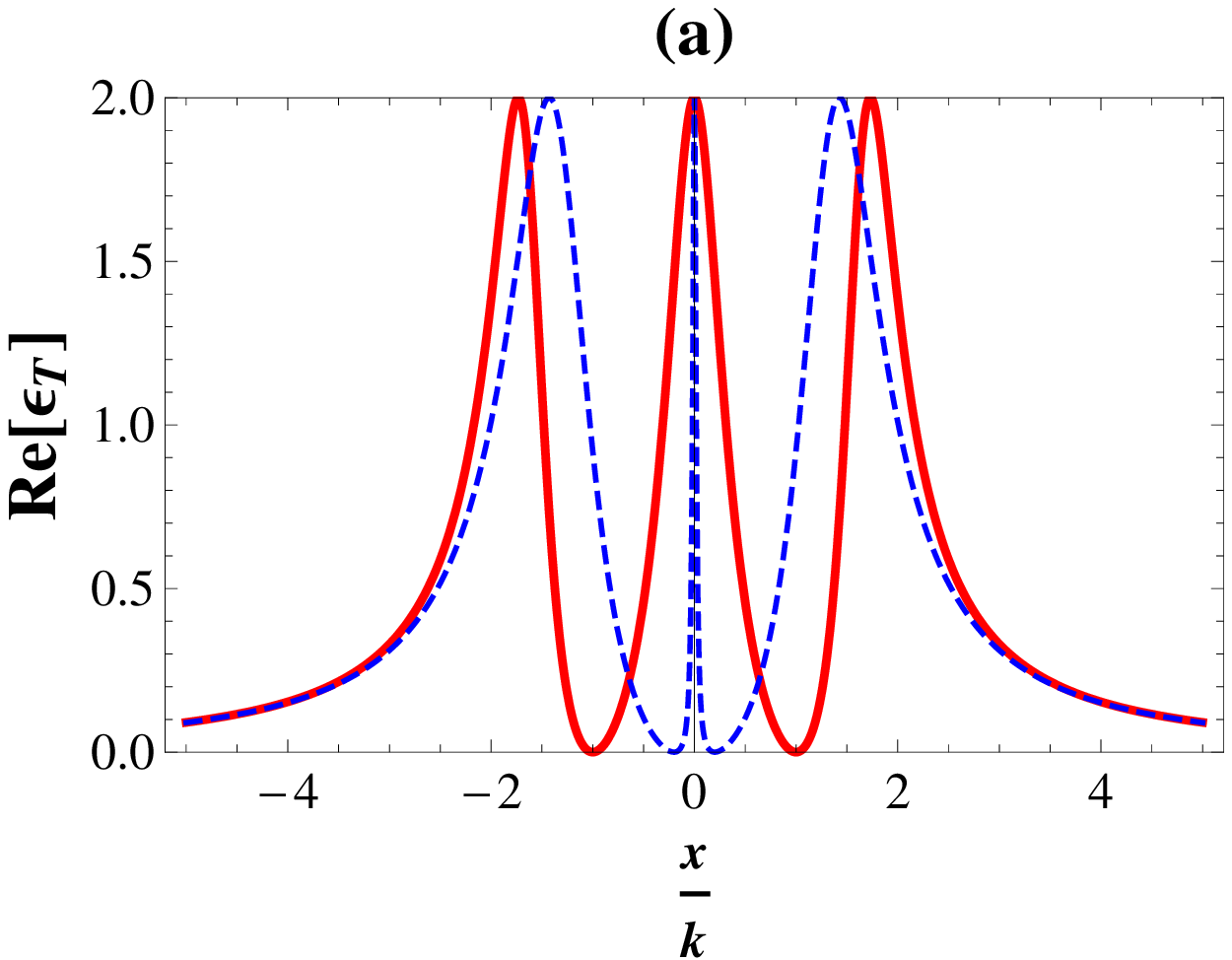}\includegraphics [scale=0.60] {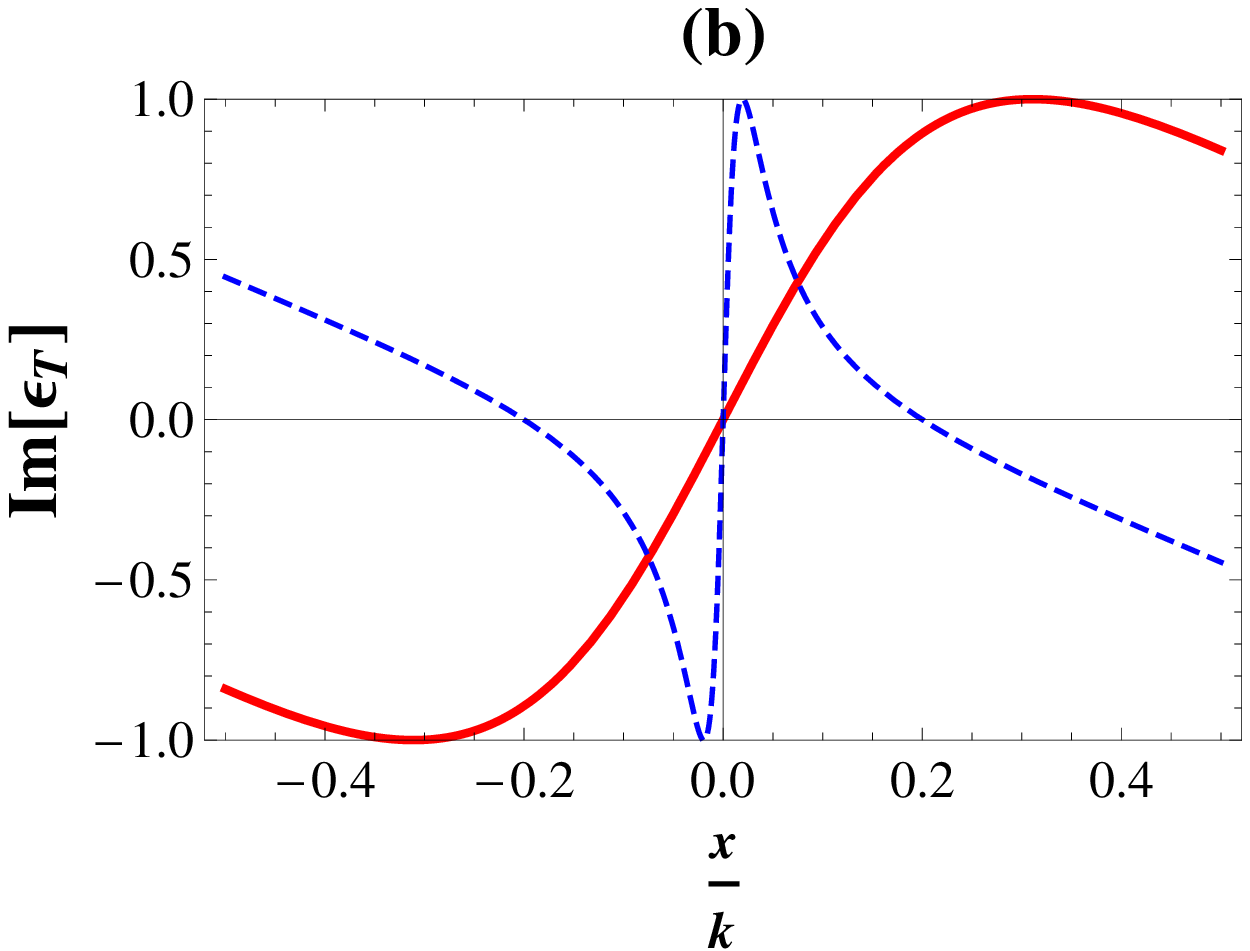}\\
\end{tabular}

\caption{(Color online)  Real (plot a) and Imaginary (plot b) part of the left-hand output probe field $\epsilon_{T}$ as a function of normalized probe detuning $x/k$. For linear case we have taken $g=k$ (solid red-line) and for the non-linear we choose $G_{N}={{0.1}{k}}$ (dashed blue-line). For all curves the other parameters are as, $G={k}$, ${\sigma^{z}}=-1$, $n=1$, ${\theta}=3 {\pi}$, $k_{d}=0$}.
\label{Fig9 (a,b)}

\end{figure}

\begin{equation}
\epsilon_{T}= {\frac{G^{2}2k ( n {\epsilon_{R}} {e^{i{\theta}}} ( - i x + {\frac{k_{d}}{2}}) + n^{2} {\epsilon_{L}} ( - i x + {\frac{k_{d}}{2}}))+ {2}{k}{\epsilon_{L}} ( - i x + k ) [ ( - i x + {\frac{\gamma_{m}}{2}})(- i x + {\frac{k_{d}}{2}}) - g^{2}  {\sigma_{z}}] }{{\epsilon_{L}}  [ ( - i x + {\frac{\gamma_{m}}{2}})(- i x + {\frac{k_{d}}{2}}) - g^{2}  {\sigma_{z}}] ( - i x + k )^{2} + (- i x + {\frac{k_{d}}{2}}) G^{2} ( n^{2} + 1)( - i x + k ){\epsilon_{L}}}}
\end{equation}
 
and for nonlinear coupling as,

\begin{equation}
{\epsilon_{T}}= {\frac{G^{2}2k ( n {\epsilon_{R}} {e^{i{\theta}}} ( - i x + {\frac{k_{d}}{2}}) + n^{2} {\epsilon_{L}} ( - i x + {\frac{k_{d}}{2}}))+ {2}{k}{\epsilon_{L}} ( - i x + k ) [ ( - i x + {\frac{\gamma_{m}}{2}})(- i x + {\frac{k_{d}}{2}}) - {4} {G_{N}}^{2} {\sigma_{z}}] }{ {\epsilon_{L}} [ ( - i x + {\frac{\gamma_{m}}{2}})(- i x + {\frac{k_{d}}{2}}) - {4} {G_{N}}^{2} {\sigma_{z}}] ( - i x + k )^{2} + (- i x + {\frac{k_{d}}{2}}) G^{2} ( n^{2} + 1)( - i x + k ){\epsilon_{L}}}}
\end{equation}

\begin{figure}[ht]
\hspace{-0.0cm}

\begin{tabular}{cc}
\includegraphics [scale=0.60] {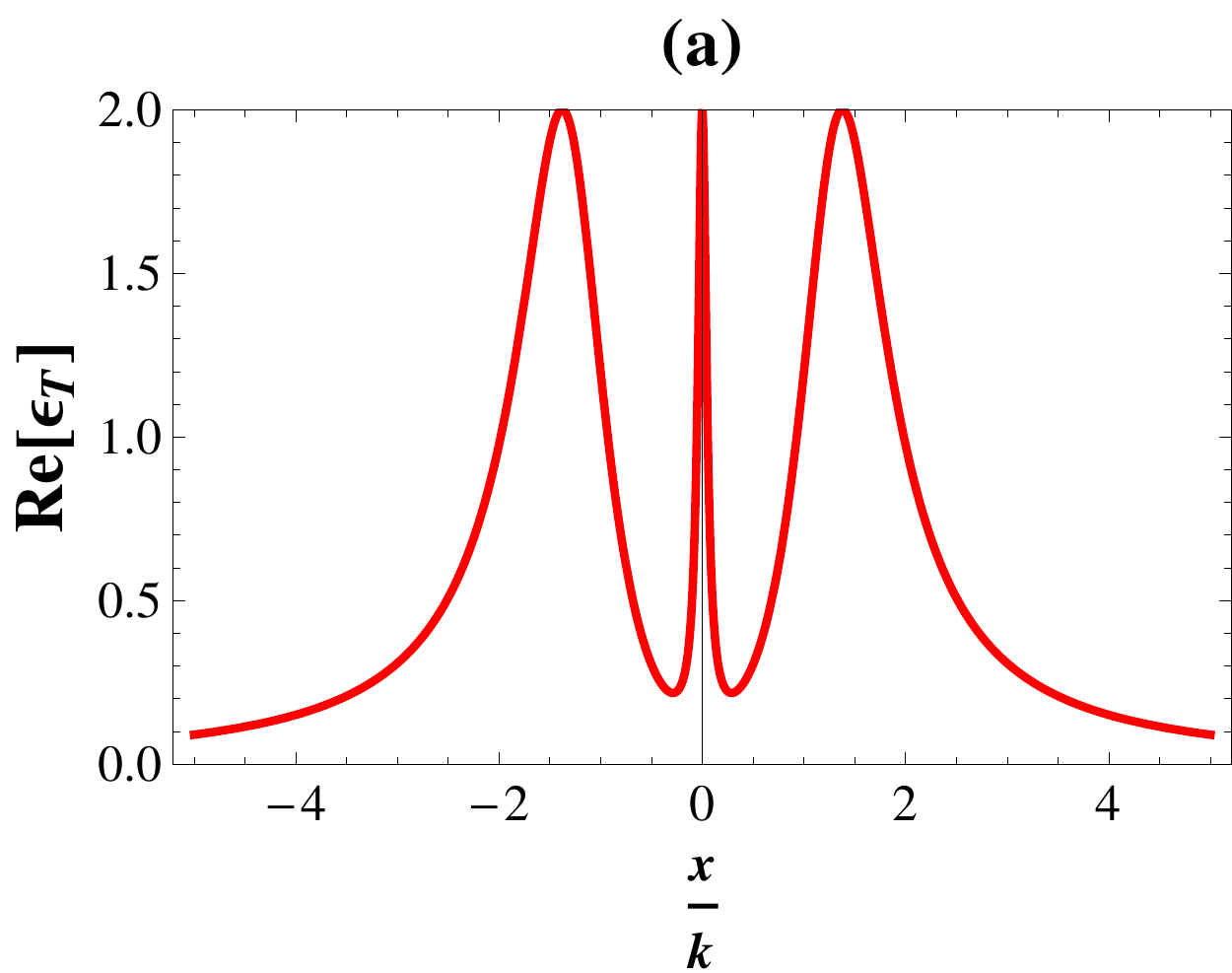}\includegraphics [scale=0.60] {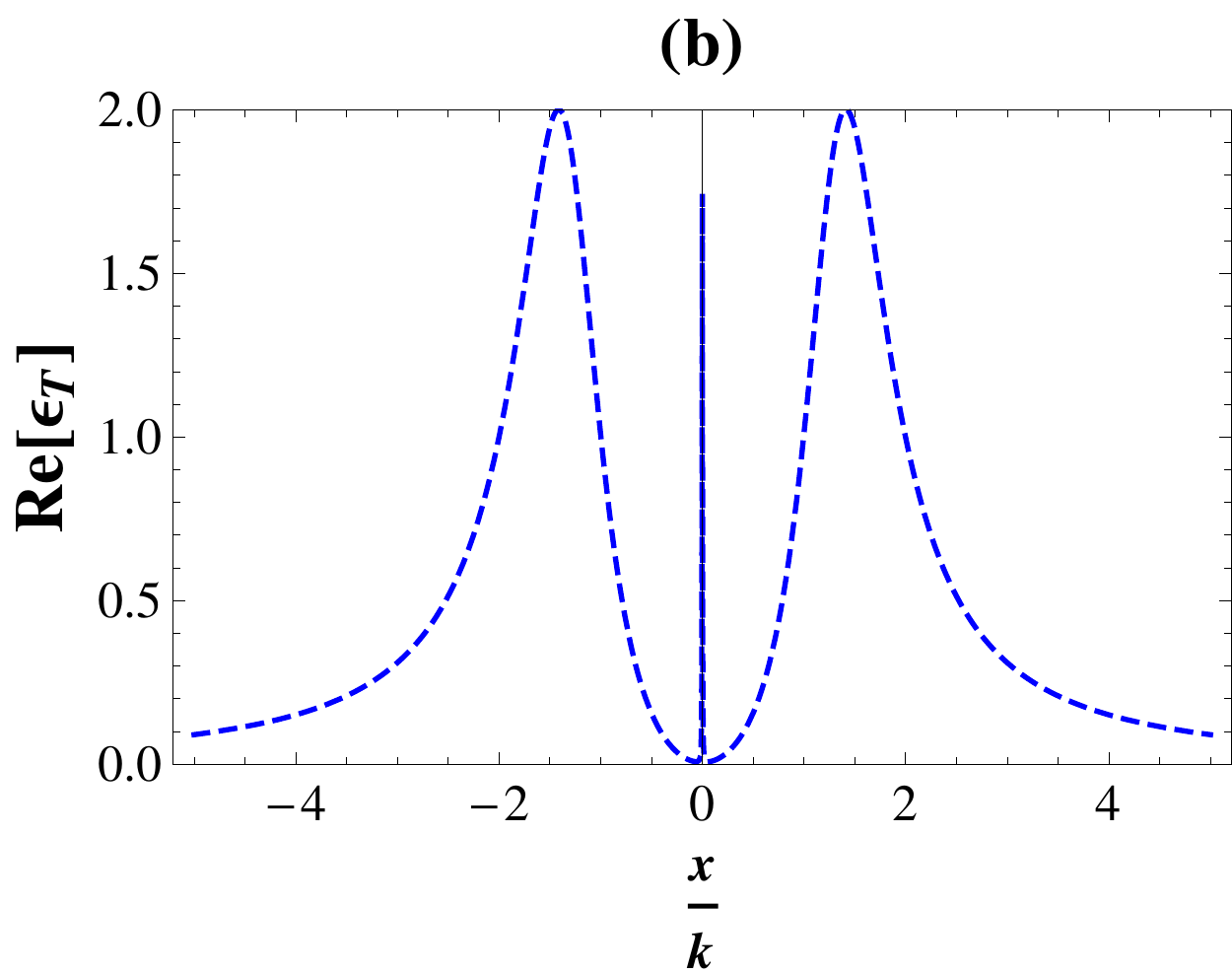}\\
\end{tabular}

\caption{(Color online) Real part of the left-hand output probe field, $\epsilon_{T}$ for the linear case (plot a) and nonlinear case (plot b) as a function of normalized probe detuning $x/k$. For linear we have taken $g=k$ (solid red-line) and for the non-linear case we have taken $G_{N}={{0.1}{k}}$ (blue dashed-line). For both the cases, the other parameters are: $G={k}$, ${\sigma^{z}}=0.1$, $n=1$, ${\theta}=3 {\pi}$, $k_{d}=0$ }.
\label{Fig10 (a,b)}

\end{figure}

Fig. 10(a) and 10(b) shows the Re$[{\epsilon_{T}}]$ and Im$[{\epsilon_{T}}]$ as a function of probe detuning $x/k$. The solid red-line depicts the linear case ($G=k$, $g=k$) and the blue dashed-line depicts the nonlinear case ($G=k$, $G_{N}=0.1k$). From Fig. 10(a), we notice that for both the linear and nonlinear case there are three absorption peaks. The width of the central absorption peak (at $x=0$) is extremely narrow for the nonlinear case as compared to the linear case. In the absence of the two level coupling with the mechanical oscillator i.e, ($g=$ $G_{N}=0$), the OMIA structure transforms to OMIT structure similar to that found in \citep{78}. Working in the mechanical red side band $\Delta_{1}=\Delta_{2}=\omega_{m}$ leads to $\Delta_{c}=\omega_{0}-\omega_{c}=\omega_{m},$ or $\omega_{0}=\omega_{c}+\omega_{m}.$ Now the central absorption peak is observed at $x=0$ i.e, $\omega_{p}-\omega_{c}=\omega_{m},$ or $\omega_{p}=\omega_{c}+\omega_{m}.$ The mechanical resonator is driven resonantly when the beat of the probe field and the control field $\delta$ matches the mechanical resonance frequency $\delta=\omega_{m}$. 

The mechanical oscillation leads to creation of sidebands of the optical field. The dominant sideband has the same frequency as the probe field. This coherent process leads to interference between the sidebands and the probe field. In the absence of the qubit, destructive interference is generated leading to cancelation of the intracavity field, resulting in a transparency window in the cavity output. On the other hand, presence of the qubit induces a sideband that is in-phase with the probe field and hence leading to a constructive interference. This results in a opaque window in the cavity. The additional absorption peaks (at $x\neq 0)$ appear due to constructive interference between higher order sidebands ( generated due to the nonlinearity in the system ) and the probe fields.

In equations (42) and (43), if we put $g=$ $G_{N}=0$, respectively and neglect $\gamma_{m}$ and $k_{d}$ compared to $k$, we obtain, 

\begin{equation}
\epsilon_{T}=\frac{ G^{2}2k ( n \frac{\epsilon_{R}}{\epsilon_{L}} {e^{i{\theta}}} + n^{2}) -2kx(x+ i k) }{ ( k - i x ) [ G^2 ( n^2 + 1) - x (x+ i k) ]}
\end{equation}

Here in eqn. (44), we find that the numerator is quadratic in x and the denominator is cubic in $x$. On the other hand in equations (42) and (43), the numerator is cubic in $x$ and the denominator is quartic in $x$. These changes determine the physical behaviour of the output field. Figure 10(b) shows the dispersion curves for the linear (red-line) and nonlinear case (blue-dashed line). Clearly the nonlinear curve is much steeper than the linear curve. The steep curve again indicates the possibility of using the nonlinear hybrid system as an optical switch.

\begin{figure}[ht]
\hspace{-0.0cm}

\begin{tabular}{cc}
\includegraphics [scale=0.60] {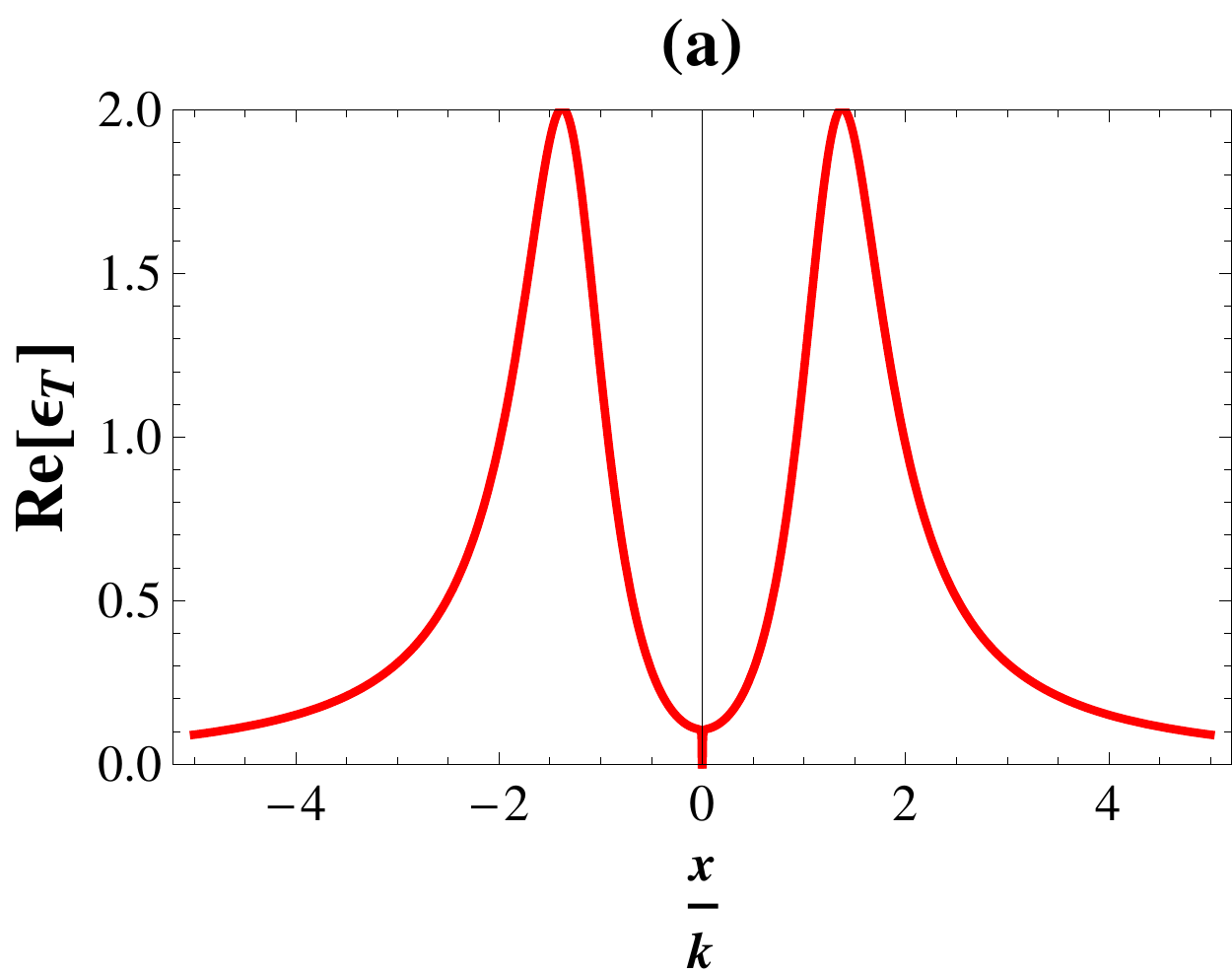}\includegraphics [scale=0.60] {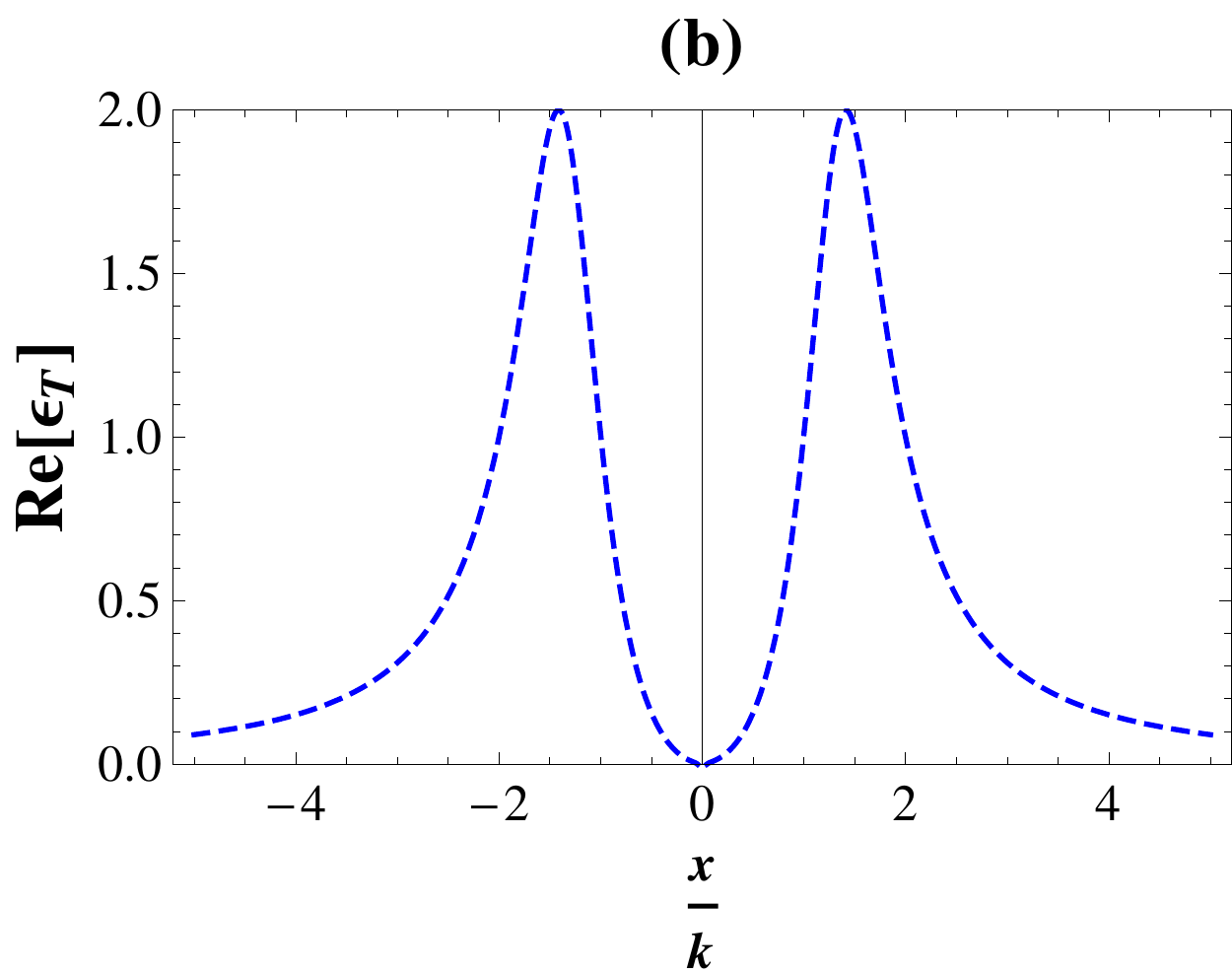}\\
\end{tabular}

\caption{(Color online)  Real part of the left-hand output probe field, $\epsilon_{T}$ for the linear case (plot a) and nonlinear case (plot b) as a function of normalized probe detuning $x/k$. For linear case we have taken $g=k$ , $k_{d}=0.1$ (solid red-line) and for the non-linear case we take $G_{N}={{0.1}{k}}$, $k_{d}=0.01$ (blue-dashed line). For both the plots, the other parameters are: $G={k}$, ${\sigma^{z}}=0.1$, $n=1$, ${\theta}=3 {\pi}$}.
\label{Fig11 (a,b)}

\end{figure}

\begin{figure}[ht]
\hspace{-0.0cm}

\begin{tabular}{cc}
\includegraphics [scale=0.60] {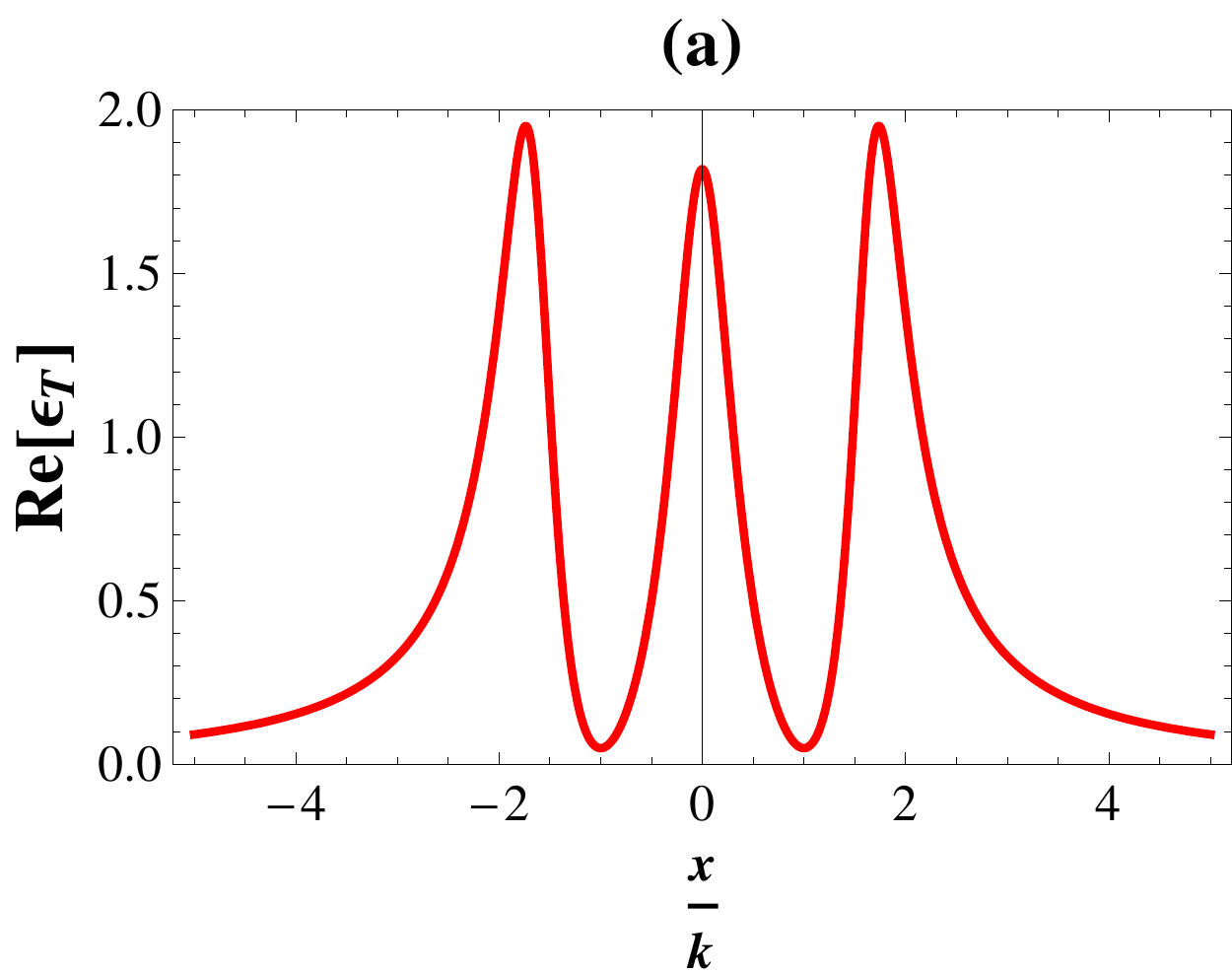}\includegraphics [scale=0.60] {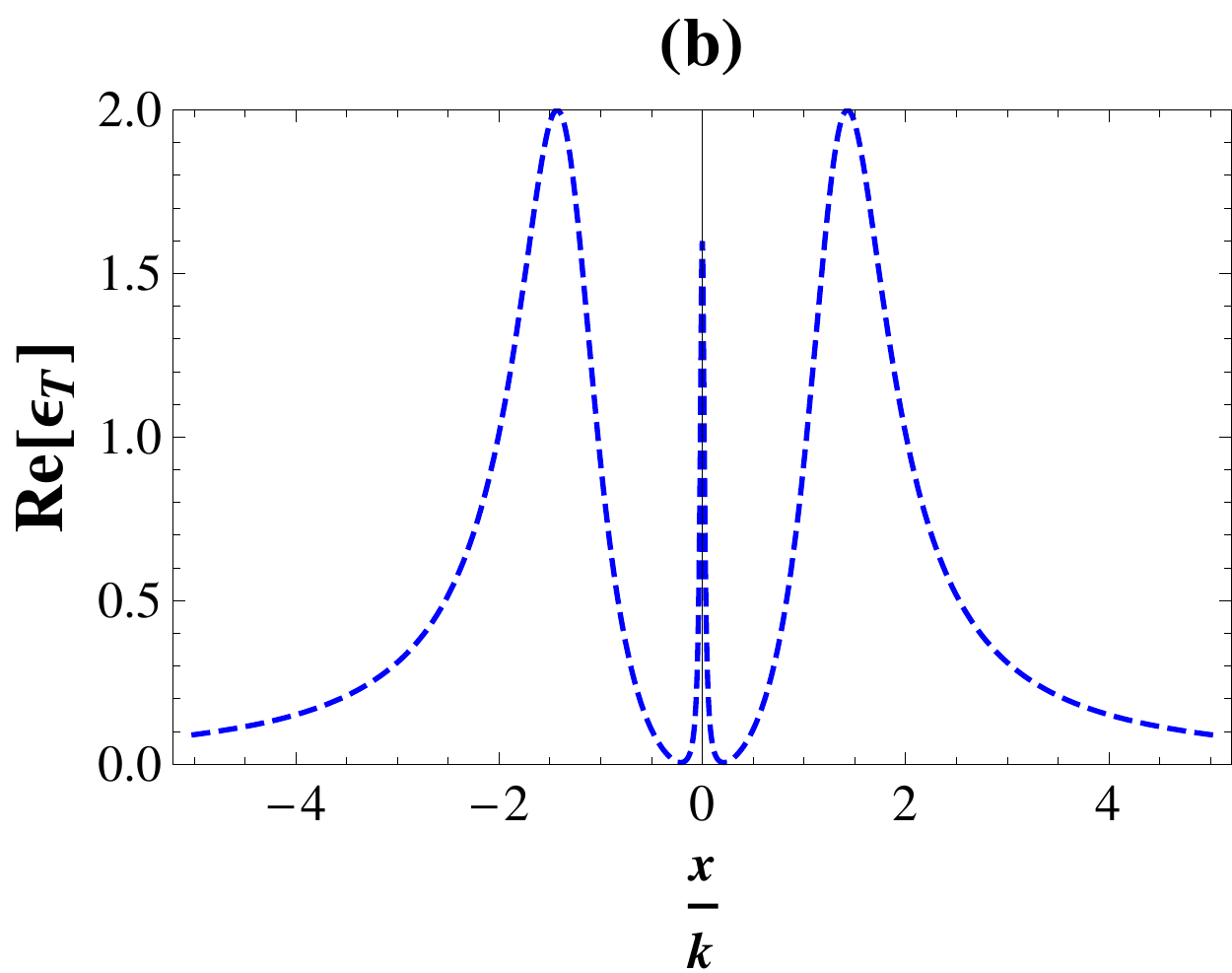}\\
\end{tabular}

\caption{(Color online) Real part of the left-hand output probe field, $\epsilon_{T}$ for the linear case (plot a) and nonlinear case (plot b) as a function of normalized probe detuning $x/k$.. For linear case we choose $g=k$ , $k_{d}=0.1$ (solid red-line) and for non-linear case we take $G_{N}={{0.1}{k}}$, $k_{d}=0.01$ (blue-dashed line). The other parameters for both the plots are:$G={k}$, ${\sigma^{z}}=-1$, $n=1$, ${\theta}=3 {\pi}$ }.
\label{Fig12 (a,b)}

\end{figure}

We now check the influence of $\sigma_{z}$ on the OMIA. If $\sigma_{z}=0$ (both upper and lower level equally populated ) then the influence of qubit-mechanical oscillator vanishes and the OMIA is converted into OMIT as in \citep{78}.  Using $\sigma_{z}=0.1$ ( population of upper level is slightly more than the lower level ), we generate the plots of Re$[\epsilon_{T}]$ for linear and nonlinear case as shown in fig. 11(a) and  11(b) respectively. For the linear case (fig.11a) the OMIA peak at $x=0$ becomes narrow and the perfect transmission around $x=0$ no longer exists. For nonlinear case ( fig.11b) the OMIA peak at $x=0$ becomes even more narrow compared to the linear case and partial transmission is observed. We also find that on increasing the qubit decay rate $k_{d}$, the constructive interference that leads to OMIA starts to disappear and a transition towards OMIT occurs. This is illustrated in figs. 12(a) and 12(b) for different values of $\sigma_{z}$ and $k_{d}$ for linear and nonlinear case. For the linear case a complete transition to OMIT occurs at $\sigma_{z}=0.1$ and $k_{d}=0.1k$. On the other hand for the nonlinear case, OMIT is seen to occur at  $\sigma_{z}=0.1$ and $k_{d}=0.01k$. For the $\sigma_{z}=-1$ case, partial OMIA is observed in figs. 13(a) and 13(b) respectively.

All the parameters used in our calculations are accessible in earlier experiments as discussed in the following \citep{24,gigan,bariani,grob,chak,reith}. The length of the optical cavity may vary from $10^{-3}-25\times10^{-3} m$. Effective mass of the mechanical mirror can vary between $5-145ng$ and its frequency varies between $1-10MHz$. The corresponding damping rate of the resonator is $\gamma_{m}=\omega_{m}/Q$, where $Q=10^7$ is the Quality factor of the optomechanical cavity. The external laser pump strength can vary from $0.2-0.5\omega_{m}$. Also, the damping rate of intracavity optical field may vary from $2\pi\times 0.1kHz-2\pi\times 1.0 MHz$. The damping rate of the two-level system may vary from $2\pi\times 0.1 MHz-2\pi\times0.66MHz$ and the linear and nonlinear coupling can be around $2\pi\times 1.0 MHz-2\pi\times 2.0 MHz$ with $g_{N}< g$\citep{73,74}. The effective optomechanical coupling $G$ can be around $2 \pi \times 2.0-3.0 MHz$.
This model can be realized experimentally by using known standard procedures. The two optically coupled cavities can be fabricated with the help of a set of distributed Bragg reflector (DBR) mirrors. Light in the x-direction can be confined by the DBRs and  the confinement along the y-z plane cab be achieved by air guiding dielectric \citep{gudat}. DBR mirror is fabricated using alternating quarter-wavelength thick high and low refractive index layers. The reflectance of DBR is dependent on the number of pairs and the difference between high and low index pairs \citep{choy}. The first and the last layers are made of AlGaAs which increases the coupling of light in/out of the structure \citep{choy}. GaAs based mechanical resonators are fabricated using a well know method micromachining  with selective etching \citep{yama,bohm}.

\section{Conclusion}
In summary, we have studied the optical response properties of a hybrid double cavity optomechanical system in the presence of a linear and nonlinear qubit-mechanical oscillator interaction. Our results illustrate that coherent perfect transmission and synthesis can be achieved at four different points which scans a wide parameter regime. From our studies it is clear that the qubit and its interaction with the mechanical mode appears as a new handle to control photon transport through the system. We further found that the case of nonlinear interaction to be more sensitive to variations in the system parameters compared to the linear case thus making it a suitable candidate for all-optical-switching. In addition, we have shown that the system exhibits opto-mechanicanically induced absorption. The system can be made to switch between OMIA and OMIT by tuning the parameters of the qubit and qubit-mechanical coupling. Thus the hybrid system can be made to operate as a tunable-photon-router as well as an all-optical-switch. Such a four-mode hybrid system with a tunable and sensitive optical response properties provides a platform for novel photonic quantum devices which can form a part of a wider quantum network.

\section{Acknowledgements}

Sabur A. Barbhuiya acknowledges BITS, Pilani Hyderabad campus for the doctorate institute fellowship.

\end{document}